\documentclass[aip,jcp,preprint]{revtex4-1}
\usepackage{fullpage}
\usepackage{graphicx}
\usepackage{setspace}
\usepackage{array}
\usepackage{color}
\usepackage{xspace}
\usepackage{amssymb}
\usepackage{amsmath}
\usepackage{mathtools}
\usepackage{multirow}
\usepackage{booktabs}
\usepackage{csquotes}
\usepackage{marvosym}

\def\m {SC$_{10}$}
\def\a {SC$_{9}$NH$_2$}
\def\c {SC$_{9}$COOH}
\def\aum {Au$_{140}$(SC$_{10}$)$_{62}$}
\def\aua {Au$_{140}$(SC$_{9}$NH$_2$)$_{62}$}
\def\auc {Au$_{140}$(SC$_{9}$COOH)$_{62}$}
\def\AA {$\mathring{A} $ }

\begin{document}
\setstretch{1.5}

\title{Gold Nanoparticles Passivated with Functionalized Alkylthiols: Simulations of
Solvation in the Infinite Dilution Limit}
\author{Saurav Prasad}
\email[Author for correspondence:]{ s.prasad@theo.chemie.tu-darmstadt.de}
\email[Tel:]{ +49-6151 16-22613}
\email[Fax:]{ +49-6151 16-22619.}
\affiliation{Eduard-Zintl-Institut f{\"u}r Anorganische und Physikalische Chemie, Technische Universit{\"a}t Darmstadt, Alarich-Weiss-Street 8, D-64287 Darmstadt, Germany.}
\author{Madhulika Gupta}
\affiliation{Department of Chemistry, Indian Institute of Technology Delhi, New Delhi:
110016, India.}
\date{\today}

\begin{abstract} 
Solvation of gold nanoparticles passivated with end group functionalized alkylthiols, namely 
	CH$_3$, NH$_2$ or COOH is
studied in solvents of varying degrees of repulsion-dispersion and
electrostatic interactions ranging from strongly polar SPC/E water to modified hybrid water models, where the Lennard-Jones contribution to
the potential energy is enhanced relative to SPC/E to completely non-polar, decane. The 
	effects due to solvent reorganization around the nanoparticle as a function of
the ligand and solvent chemistry are monitored using the nanoparticle-solvent
pair correlation functions and tetrahedral order parameter. The solvent penetration inside the ligand shell is maximum
for decane, indicating better solute-solvent interaction in decane compared to other solvents. The 
	COOH end group functionalized nanoparticle breaks the
tetrahedral structure of water molecules more as compared to other nanoparticles
used in this study. The ligand reorganization and its effect on solvation are
monitored using radial density profiles (RDPs), radius of gyration ($R_g$) and 
ligand asymmetry parameter ($\langle \Delta \rangle$). RDP and $R_g$ values 
	show significant stretching of ligands in decane
	than in model waters, which is also consistent with $\langle \Delta \rangle$. 
	The ligand shell anisotropy for all nanoparticles is maximum in SPC/E water and minimum in
decane. The isotropic 
potential of mean force ($V_{PMF}(r)$) between two identical end group functionalized
nanoparticles have been calculated in vacuum, SPC/E water and H3.00 modified hybrid water, which consistently shows 
	attractive well depth. Distance-dependent
fluctuation driven anisotropy has also been examined. The implications
for self-assembly of passivated gold nanoparticles from aqueous dispersions as well as the dependence of 
calculated quantities on ligand and solvent chemistry are highlighted.
\end{abstract}
\maketitle

\section{Introduction}

Inorganic nano-cores passivated with different ligands with charged or 
uncharged tails self-organize into unique complex structures and show
various physical, chemical and biological phenomena, which also depend on size and shape of the passivated
nanoparticle\cite{npk10,da04,bwsg09,wkcg11,ulm01,tpssl94,ber06,fflm08,wzrf02,eclyt01,wxc99}. The advantage of these
size-dependent properties can be further magnified and tailored by obtaining self-assembly of these
nanoparticles into supraparticular assemblies of varying dimensions and
superlattices\cite{da04,bwsg09,tkg02,kfbbs98,mkb00,mkb95,stkom06,tsbycm09,lsm99,tzwgk06,zewm06}. In order to obtain an ordered
2D or 3D super-structure from self-assembly process, it is very important to
get a stable dispersion of that particular nanoparticle (NP) in a suitable solvent.
The passivation of the nano-core with suitable ligands leads to the formation of stable
nanoparticle dispersions in various solvent media and is the initial step in order
to obtain bulk nano-structures due to NPs self-organizing
ability. The interaction between NPs and the subsequent formation of self-organized
nano-structures not only depends on the shape, size and type of the nano-core but also on the
ligands used for passivation and the solvent medium\cite{makgi08,bwsg09}, since the solvation free
energy also takes into account the ligand-ligand and ligand-solvent
interactions.  Hence, understanding the
behaviour of different ligand passivated nanoparticles in various solvents and
the reorganization of the solvent around the passivated nanoparticle is of prime importance in order to elucidate the
reasons for the formation of unique structures due to self-assembly process and their relation to several phenomena. 

In order to get a passivated nanoparticle, various combinations of nano-core and
coatings have been reported in the literature\cite{npk10,da04,pjkjh07,bwsg09,wkcg11}. The nano-core can be metal or its
oxide, solid polymers, semiconductors; whereas coatings can be of a thin layer of a hard material on another
material to obtain nanoshell particles or can be soft ligands such as small
organic molecules, polymers or biological
materials. The $Au$ core 
and the alkanethiol ligands have been extensively studied and the potential energy
functions are well established to model such
systems\cite{da04,lg10,ll98,drmsw10}. The significance of this system in
nanofiltration, drug delivery, biochemical
            sensors and optoelectronics and other fields have been well illustrated in the
            literature\cite{hldj11,hldj11nano,esmlm97,ba09,ghmkr08}. In the
present study, we have studied nanoparticles composed of $Au$
nano-core uniformly passivated with various end group functionalized
alkylthiols.

The basic interactions governing the self-assembly process include van der
Waals, magnetic, molecular surface forces, entropic effects and electrostatic
interactions\cite{bwsg09}. Open structures can be obtained from self-assembly process due to
electrostatic interactions, whereas, all other interactions or effects mostly
form closed packed structures\cite{bwsg09}. The nanoscale self-assembly due to
electrostatic interactions can be motivated by appropriate surface
functionalization or $\omega$-functionalization of ligands used to passivate
bare nano-cores. Experimentally, self-assembly of binary nanoparticles of almost
equal size, passivated with oppositely charged $\omega$-functionalized
alkylthiols are shown to form large 3D-crystals with diamond-like lattice\cite{kfpsbg06,bcg13}. Each NP is surrounded
by four oppositely charged neighbors at the vertices of the tetrahedron. Initial
molecular dynamics studies of self-assembly of alkanethiol passivated gold
nanoparticles have been done by Luedtke and Landman\cite{ll96,ll98}. Subsequently, several
studies have been performed to understand various aspects of alkanethiol coated
gold nanoparticles\cite{tb05,tb06,rz07,gg07,sghjlsg07,spv08,drsm13,lg14} followed by recent computational studies of
$\omega$- or end group functionalized nanoparticles.
Yang and Weng studied the structure and dynamics of water, when nanoparticles passivated with different neutral end group
ligands are solvated in water\cite{ywc11}. Lane and
Grest have also studied similar systems, focusing mainly on the spontaneous
asymmetry of the passivated nanoparticles, which seem to depend on the chain
length, particle size and thermodynamic variables like temperature and have
significant implications on self-assembly process\cite{lg10}. They have also further extended their
study to understand the effects of charged ligands on coating
asymmetry\cite{blg14}. Henz et al. have concentrated on determining the
binding energy, density and solubility parameters of functionalized gold
nanoparticles in vacuum\cite{hcaclb11}.
Heikkila et al. have studied charged mono-layer $Au$ NPs with particular emphasis on 
electrostatic properties\cite{hgmhva12}. Giri and Spohr have also studied nanoparticles
passivated with neutral and charged end group ligands with varying grafting
densities of the ligands, with special attention on
the penetration of water and ions in the soft corona\cite{gs15}. Lehn et al. have studied
mixed-mono-layer protected gold nanoparticles, where only one of the ligands is end
functionalized\cite{la13}.

In this study, we have carried out simulations of three types of $\omega$- or end group functionalized 
nanoparticles. The
structure, thermodynamics and dynamics of a particular solvent also play an
important role in the self-assembly process. Since the solvent properties can be easily varied either by changing
the density or temperature, they provide an easy way to control the self-assembly
process. Solvents are chosen to have varying degrees of repulsion-dispersion, electrostatic interactions and simple liquid character. The solvents considered in this study
are SPC/E water model, modified hybrid water models
(H1.56 and H3.00) and decane. SPC/E water is anomalous and polar in nature and has both
repulsion-dispersion and electrostatic interactions\cite{bgs87}. For modified
hybrid water models
(H1.56 and H3.00), the degree of repulsion-dispersion contributions to the
potential energy is increased by a factor of 1.56 and 3.00 with respect to
SPC/E\cite{ld05}, whereas decane is non-polar\cite{ms98}. H1.56 and H3.00 are chosen, since they have been 
extensively studied by their developers as well as in our group\cite{ld05,lh06,al07,cdsl08,lb10,pc14}. The modified hybrid water
models were designed with a view to understand the relative contribution of LJ dispersion-repulsion 
term to the electrostatic term in determining the bulk and solvent properties of water. As the weight of the LJ 
term relative to the electrostatic term increases, a set of liquids which may be regarded as hybrids between SPC/E water 
and LJ liquid are created. Therefore, the modified hybrid water models may also be taken as representative of a range
of strong and moderately polar liquids\cite{pc14}. This particular study
has been carried out to examine
the effect of solvent interactions, structure and dynamics of the solvent and also the effect of end group
functionalization on the thermodynamics of solvation, coating asymmetry and its implications on the
self-assembly process.

The remaining paper has been organized as follows.
Section~\ref{sec:obser6} describes the various observables calculated in this study.
Section~\ref{sec:comput6} discusses the potential energy surface and the
molecular dynamics simulation details. The results and conclusions are given in
Sections~\ref{sec:result6} and~\ref{sec:conclu6}, respectively.
 
\section{Solvation Structure and Thermodynamics}  
\label{sec:obser6}
This section discusses the ideas which we have used to quantify the solvation
structure and relate it to the thermodynamics of solvation of a single passivated
nanoparticle in water, modified hybrid water models (H1.56 and H3.00) and decane. We have used Ben-Naim 
approach in order to understand the
solvation behaviour\cite{abn78,nyjc12}. In all our single nanoparticle
simulations at constant volume and temperature, the
rigid gold-core is held fixed, keeping the center of mass of the gold-core fixed
at the center of the solvent box, which eliminates the translational and
rotational motions of the gold nano-core. The ligand and solvent molecules are
free to reorganize themselves under the influence of each other. The pair correlation
function, $g_{ns}(r)$, between the center of mass of the gold nano-core and the
oxygen atoms of water (water as solvent) and decane monomer units (decane as solvent) describes the reorganization of solvent molecules around the
ligand passivated nanoparticle. The pair correlation function, $g_{ns}(r)$, acts
as central structural property, which can be used to define other important
solvation properties like solvent excess ($n^{E}_{s}$), 
the local ($\Delta S^{loc}_{ns}$) and
long-range ($\Delta S^{long}_{ns}$) contribution to entropy 
to quantify the solvation behaviour of individual nanoparticles.

The solvent excess, $n^{E}_{s}$, is defined in terms of the Kirkwood-Buff integral
as\cite{kb51,lps06}
\begin{equation}
  n^{E}_{s}= {4\pi \rho_{i}} \int^{\infty}_{0} (g_{ns}(r)-1) r^{2} dr
  \label{eqn:nexc}
\end{equation}
where $\rho_{i}$ is the number density of oxygen atoms or decane
monomer units depending on the solvent used. $n^{E}_{s}$ acts as a quantitative measure of the
affinity of the solute for the given solvent by estimating the excess number of
solvent molecules around an infinitely dilute solute molecule\cite{pd89,dm89}. 

Since the rigid gold core is held fixed, the entropy will mostly be associated
with the reorganization of the ligand and solvent molecules under the influence
of each other. 

\subsection{Entropy of the Ligand Shell} 
The upper bound to the entropy of the
flexible, long-chain ligand corona called the ligand shell configurational entropy,
$S_L$, can be estimated by the Schlitter's method using the covariances of
the ligand atomic coordinates using the formula\cite{js93},
\begin{equation}
  S_{abs} < S_L= \frac{1}{2}k_B \ln \left|{\bf{1}}+\frac{k_BT
  e^2}{\hbar^2}\bf{M}^{1/2}\sigma \bf{M}^{1/2}\right|
  \label{eqn:sl}
\end{equation}
where, $k_B$ is the Boltzmann constant, $e$ is the Euler number and is equal to
$\exp(1)$, $\bf{M}$ is the 3N-dimensional diagonal matrix having masses of $N$
atoms of corresponding chains, $T$ is the simulation temperature and
$\sigma$ is the covariance of the ligand atomic coordinates, defined as
\begin{equation}
  \left<\sigma_{ij}\right>= \left<(x_i -\left<x_i\right>)(x_j-
  \left<x_j\right>)\right>
  \label{eqn:covar}
\end{equation}
where, $x_i$ and $x_j$ are the Cartesian coordinates of $i^{th}$ and $j^{th}$
atoms. Since, the ligand shell configurational entropy, $S_L$, is the upper bound to entropy, it 
will always be greater than the absolute entropy of the system, $S_{abs}$. 
The entropy is calculated by using coordinates of ligand atoms after fitting each 
configuration with respect to the first frame of the trajectory in order to eliminate the translational and rotational motions of the ligand.
\subsection{Entropy due to Solvent Reorganization} 
The change in entropy due to
  solvent reorganization around the passivated
    nanoparticle is estimated using nanoparticle-solvent pair correlation function. The total entropy associated with 
  the reorganization of the solvent molecules around the ligand passivated nanoparticle is 
  given as the sum of entropies arising due to the local ordering of solvent molecules
  and a long-range correction term to the entropy of solvation
  as\cite{tl981,tl982}
\begin{equation}
  \Delta S^{tot}_{ns} = \Delta S^{loc}_{ns} + \Delta S^{long}_{ns}
\end{equation}
The local entropic contribution due to solvent ordering around a spherically
symmetric solute, $\Delta S^{loc}_{ns}$, can be written in terms of 
$g_{ns}(r)$ as
\begin{equation}
  \Delta S_{ns}^{loc}/k = - 4\pi \rho_{i} \int^{\infty}_{0}(g_{ns}(r) \ln
  g_{ns}(r))r^2 dr
   \label{eqn:sloc}
\end{equation}
and the long-range correction term is defined in terms of Kirkwood-Buff integral as
\begin{equation}
    \Delta S_{ns}^{long}/k = 4\pi \rho_{i} \int^{\infty}_{0}(g_{ns}(r)-1)r^2 dr
\label{eqn:slong}
\end{equation}

\begin{figure}[h!]
\begin{center}
  \includegraphics[clip=true,height=5.5cm,width=8.5cm]{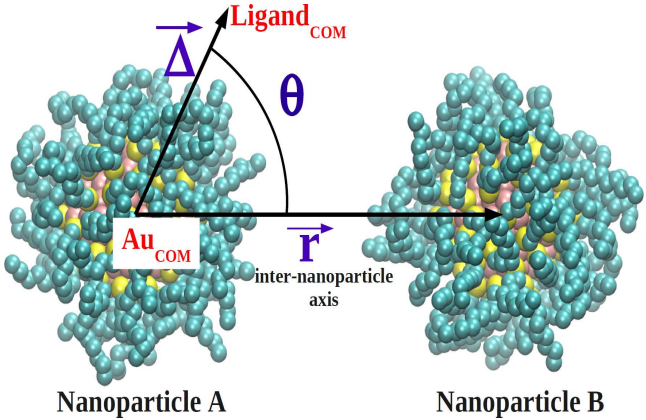}
    \end{center}
    \caption{Schematic diagram illustrating the mass dipole vector, $\vec{\Delta}$, and the 
    orientation of the mass dipole vector with respect to
      the inter-nanoparticle separation vector ($\vec{r}$). The snapshot is for
    two \aum~nanoparticles in vacuum at 300 K for the pair separation
    of 55 \AA. The sulfur atoms are shown in yellow and carbon in cyan.
    $Au_{COM}$ and $Ligand_{COM}$ represent the center of mass of the gold
    nano-core and ligand shell, respectively. The mass dipole vector,
    $\vec{\Delta}$, is defined as the vector connecting the $Au_{COM}$ and
    $Ligand_{COM}$. $\theta$ is the angle between the $\vec{\Delta}$ and the
    inter-nanoparticle axis ($\vec{r}$), joining the center of masses of the two
  nanoparticles.}
        \label{fig:delta-angle}
\end{figure}

The fluctuations of the soft ligand corona around
the passivated nanoparticle give rise to a form
of anisotropy, which is an inherent property of single nanoparticles with small
size and high surface curvature\cite{lg10,silg14,bmkcc13}. This inherent anisotropy is quantified by
calculating the mass dipole vector, $\vec{\Delta}$, between the center of mass
of the gold nano-core and
the ligand corona as shown in Figure~\ref{fig:delta-angle} and also the probability distribution of the mass dipole
vector, $P(\vec{\Delta})$\cite{bmkcc13,jykc14,ysac16}. We have used $\vec{\Delta}$ and $P(\vec{\Delta})$
to estimate the degree of inherent anisotropy.
 
\subsection{Potentials of Mean Force}
Using the simulations of a pair of identical nanoparticles in different solvent media, we
have computed the isotropic potential of mean force and an unusual form of
anisotropy called emergent anisotropy, which depends on the pair separation
between the two nanoparticles.

Consider two nanoparticles A and B with position vectors ${\bf{r}_A}$ and
${\bf{r}_B}$ as shown in Figure~\ref{fig:delta-angle}. The free energy change as a function of the pair separation,
$r=|\bf{r}_A-\bf{r}_B|$, gives the conventional isotropic potential of mean force (PMF),
$V_{PMF}(r)$. We have used constrained molecular dynamics
technique\cite{jykc14,tkg07,lctw07,spv08,pe07,tb05} to calculate
the ensemble-averaged mean force at a pair separation $r$ between two nanoparticles A and B, which can be written as the derivative of the potential energy
function, $U(r_A,r_B,r^N)$ with respect to the pair separation
$r=|\bf{r}_A-\bf{r}_B|$ as\cite{jykc14}

\begin{eqnarray}
  \left \langle \frac{\partial U}{\partial r} \right\rangle 
  = \frac{1}{2}\left[{\bf F}_A-{\bf F}_B\right]\cdot \hat{\bf r}
  \quad {\rm if}\quad m_A = m_B
\end{eqnarray}
where ${ \bf F}_A$ and ${ \bf F}_b$ are the forces acting on nanoparticles A and
B due to all solvent
molecules and $\hat{\bf r}$ is the unit vector in the direction $r=|\bf{r}_A-\bf{r}_B|$.
Then the isotropic potential of mean force (PMF) can be obtained by integrating
over a set of pair separations as 
\begin{eqnarray}
  V_{PMF}(r)=\int_{\infty}^{r} \left \langle \frac{\partial U}{\partial r}
  \right\rangle dr - 2k_BT ln\left(r\right)
\end{eqnarray}
where the last term is important when hard constraints are employed in MD
simulations\cite{tkg07,lctw07}.

In case of two nanoparticles, the mass dipole vector ($\vec{\Delta}$) can
be influenced by the presence of another approaching nanoparticle. The angle
($\theta$) between the mass dipole vector and the pair separation vector ($\bf{r}$) also provides
relevant information about the orientation of the ligand shell or mass dipole as
shown in Figure~\ref{fig:delta-angle}.
Hence, the changes in the magnitude and orientation of the mass dipole vector in presence of an approaching second nanoparticle give rise to
emergent anisotropy. Due to the presence of very high emergent anisotropy, the
isotropic PMF becomes insufficient and does not provide reasonable explanations
for many aspects of self-assembly process and hence, a two-dimensional anisotropic PMF,
$V_{anis}(r,\cos \theta)$ is required\cite{bmkcc13}. In this paper, we compute the isotropic
PMF between two identical end group functionalized ligand passivated
nanoparticles in various solvent media and characterize the
emergent anisotropy by calculating the change in magnitude and orientation of
the mass dipole vector as a function of the pair separation between the two identical
nanoparticles. 
\section{Computational Details}
\label{sec:comput6}
This section gives the details of the potential energy surface and molecular
dynamics simulation protocols used for $\omega$-functionalized ligand passivated gold nanoparticles 
in water, modified hybrid waters and decane. 
Earlier studies in
order to validate the simulation protocols have already been done in our group
for $Au_{140}$ passivated with alkanethiols in organic
solvents\cite{ysac16,yac17}. 
Subsections~\ref{subsec:pes}
and~\ref{subsec:md} describe the potential energy surface and molecular dynamics
simulation details used in our studies.
\subsection{Potential Energy Surface}
\label{subsec:pes}
  \subsubsection{Modified Hybrid Water Models}
  The modified hybrid water models were designed with a view to understanding
  the relative contribution of the Lennard-Jones dispersion-repulsion term to
  the electrostatic term in determining the bulk and solvent properties of
  water\cite{ld05,lh06,al07,cdsl08,lb10}. The potential energy form for the
  modified hybrid water models is written as
  \begin{equation} 
    U = \lambda U_{LJ} + U_{ES}
    \label{eqn:potMW}
  \end{equation}  
  where $U_{LJ}$ is the Lennard-Jones contribution and $U_{ES}$ is the
  electrostatic contribution. For the SPC/E model, the hybrid potential
  parameter, $\lambda$, is set to unity. The molecular geometry and partial charge
  distribution are constant for all values of $\lambda$. Increasing  $\lambda $
  increases the weight of the Lennard-Jones term relative to the
  electrostatic term, creating a set of fluids which may be regarded as hybrids
  between SPC/E water and a Lennard-Jones liquid. Therefore, the modified hybrid   
  water models may also be taken as the representative of a range of strong and moderately polar liquids.
 The Lennard-Jones site is located on the oxygen atom of each water molecule and
 is characterized by the energy, $\epsilon_{OO}$, and size, $\sigma_{OO}$,
 parameters. The Lennard-Jones contribution to the potential energy for a pair
 of water molecules is given by
 \begin{equation} 
    U_{LJ}\left(r\right) =
    4\epsilon_{OO}\left[\left(\frac{\sigma_{OO}}{r}\right)^{12} -
    \left(\frac{\sigma_{OO}}{r}\right)^{6}\right]
    \label{eqn:lj}
  \end{equation} 
  The electrostatic interaction between two water molecules $a$ and $b$ is given
  by
  \begin{equation}
    U_{ES}\left(r\right) = \sum_i\sum_j\frac{q_iq_j}{4\pi\epsilon_{0}r}
    \label{eqn:es} 
  \end{equation} 
  where $i$ and $j$ index the partial charges located on molecules $a$ and $b$.
  The parameters for the modified hybrid water models are summarized in
  Table~SI of the supplementary material\cite{supple-nano}.
\subsubsection{Decane, Gold-core and Ligands}
  A rigid $Au_{140}$ truncated gold cluster was used in all our
  simulations\cite{kt05}. Gold nano-cores of 140 $Au$ atoms were passivated with $\omega$-functionalized
alkylthiols (SRX), where S, R and X represent thiol group (SH), common
alkane chain (R = C$_9$H$_{18}$) and terminal group (X= CH$_3$, NH$_2$ and COOH), respectively. 
The TraPPE-UA model potential\cite{ms98}
was used for the thiol (SH), common
alkane chain (R = C$_9$H$_{18}$), the terminal CH$_3$ and decane solvent,
whereas all the atoms in NH$_2$ and COOH are simulated explicitly. Since, united atom model has been used for alkane chain 
(R = C$_9$H$_{18}$) and the terminal CH$_3$, so only symbol $C$ will be used further to denote a methyl or methylene unit. Hence, the different type of ligands 
used in this study can be written as \m, \a~and \c. All
non-bonded interactions were modeled via the Lennard-Jones and Coulomb
potential. The parameters for
similar atom pairs\cite{ms98,tb05,ll04,rj99,wsrs05,kcp04,cbr06} are given in the
Table~SII of the supplementary material\cite{supple-nano}. Lorentz-Berthelot
combination rules were applied for dissimilar atom pairs\cite{arl96}. 
Bonds and
bond angles were represented using harmonic stretching potential and harmonic bending
potential, respectively. For 1-4 torsional interactions, we have used triple cosine 
and multi-harmonic potential. The parameters for harmonic stretching, harmonic
bending, triple cosine and multi-harmonic potential\cite{ms98,tb05,ll04,rj99,wsrs05,kcp04,cbr06} are given in Tables~SIII,~SIV,~SV and~SVI, respectively of the supplementary material\cite{supple-nano}. 
We would like to
mention that we have used harmonic stretching for bonds instead 
of rigid bonds as used in original TraPPE-UA potential, both of which give 
similar results\cite{ysac16}. Additionally, the torsional equations 
related to terminal groups like NH$_2$ and COOH used in the
literature are not present in LAMMPS MD package, hence, we have fitted these
equations to obtain the constants for multi-harmonic potential. 
The fitted plots
 to obtain the constants for multi-harmonic potential are shown 
 in Figures~S1 and~S2 of the supplementary material\cite{supple-nano}. Non-bonded Morse
potential was used for the
gold-thiolate interaction; $U(r)=D_e((1-exp(-k(r-r_e)))^2-1)$, where, $r_e$ = 2.9 \AA~is the equilibrium distance, $D_e$ = 38.6 kJ mol$^{-1}$ is the well depth, and $k$ = 1.3 {\AA}$^{-1}$ controls
the range of the potential\cite{ll98}. 

\subsection{Molecular Dynamics Simulation Details}
\label{subsec:md}
We report molecular dynamics simulations of: (a) bulk water and decane,
(b) single $\omega$-functionalized ligand passivated nanoparticle in vacuum, water and decane
and (c) a pair of identical $\omega$-functionalized ligand passivated nanoparticles in
vacuum and water. All simulations were performed using LAMMPS package with GPU
acceleration\cite{lammps}. A timestep of 1 fs was used to integrate the equations of
motion using the velocity-Verlet algorithm. The simulation temperature and pressure
were maintained, wherever applicable, using Nose-Hoover thermostat and barostat
with a relaxation time of 0.1 ps and 1 ps, respectively. Rigid body
constraints in water were maintained using the SHAKE algorithm with tolerance value of
10$^{-6}$ \AA. PPPM was used to account for the long-range electrostatic
interactions. All non-bonded
interactions were truncated at 14 \AA~during the solvation studies of passivated
nanoparticles and in vacuum, a spherical cutoff of 30 \AA~was used.
\subsubsection{Passivation of $Au _{140}$ Core in Vacuum}
We followed the protocol similar to Luedtke and Landman for passivation of the gold
core with all types of $\omega$-functionalized ligands\cite{ll96,ll98}.
The $Au_{140}$ core was placed
at the center of a 100 \AA~$\times$ 100 \AA~$\times$ 100 \AA~cubic cell surrounded randomly with
100 decanethiol chains using PACKMOL package\cite{mabm09}. The system was equilibrated in NVT
ensemble for 1 ns at 200 K in vacuum keeping the gold
nano-core stationary. To ensure equilibration, rescaling of velocities
were performed at every 10 steps. The system obtained after initial equilibration
was heated from 200 to 500 K using a ramp of 2.5 K/ps. The system was then
gradually cooled back from 500 to 300 K using a ramp of 0.5 K/ps. After the end of
passivation protocol, 62 ligands were found attached to the gold core, consistent with
the number of binding sites available. The number of attached ligands are also consistent 
with previous publications\cite{ll96,tb05}. All other excess ligand chains were removed.
To ensure the stability of the passivated gold nanoparticle, 2 ns simulation at 300
K was performed. For passivation of the gold core using SC$_9$NH$_2$ and SC$_9$COOH ligand chains,
the same procedure was followed, except that the system obtained after initial
equilibration of 1 ns at 200 K was heated from 200 to 500 K using a ramp of 0.15
K/ps. Production runs of 20 ns were performed for all three types of passivated
(\aum, \aua, \auc) nanoparticles in vacuum.
\subsubsection{Solvation of a Passivated Nanoparticle}
We have used water (SPC/E or H1.00), modified hybrid water (H1.56 and H3.00) and
decane as a solvent. Solvent box for water was prepared by randomly packing the
water molecules in 100 \AA~$\times$ 100 \AA~$\times$ 100 \AA~cubic simulation cell with density
1.00 g/cc (33428 water molecules). For decane, 0.73 g/cc density was used and
decane molecules were also randomly packed in a cubic box
of same dimensions (3090 decane molecules). 
The passivated nanoparticle 
was inserted in the solvent box by growing a void of
18 \AA~in radius in the center of the cubic box by using a soft repulsive
spherical indent as shown in Figure~S3 of the supplementary material\cite{supple-nano}. After insertion, energy minimization of the system was done
using conjugate gradient method in NVT ensemble for 1 ns. 
For all the solvated passivated nanoparticle systems prepared,
equilibration runs of 10 ns were carried out in the NPT ensemble at 275 K and 300 K with 1 atm pressure. Once an equilibrium volume was obtained, the barostat
was switched off and NVT production runs for 20 ns were performed at the same
temperatures.

\subsubsection{Potentials of Mean Force in Vacuum and Water}
  In vacuum, the two passivated nanoparticles were placed along the x-axis at a
  separation of 55 \AA~between the center of masses of the two nano cores
  in an orthorhombic simulation cell of 160 \AA~$\times$ 100 \AA~$\times$ 100 \AA.
  The gold core was held fixed in all the simulations. At this separation,
  equilibration run of 4 ns and production run of 5 ns were performed with data
  sampling frequency of 10 steps to generate ensemble averaged mean forces
   at 300 K and 1 atm. The pair separation between the nanoparticles 
  was reduced along the x-axis by keeping one nanoparticle fixed and moving the
  center of mass of the other nano core very slowly (10$^{-6}$ \AA~per time steps)
  towards the fixed nanoparticle. Similar lengths of equilibration and production runs were
  carried at each pair separation up to 24 \AA~separation with
  the sequential decrease of 1 \AA. In a solvent box of dimensions 160 \AA~ $\times$
  100 \AA~ $\times$ 100 \AA (1.0 g/cc water, 54078 water molecules), two voids were generated at 55
  \AA~separation and passivated nanoparticles were placed inside these voids using VMD package
  as done for single nanoparticles\cite{vmd96}.
  At this pair separation, equilibration and production runs of 4 ns and 5 ns, respectively, were performed. Then the separation
  was reduced as mentioned above from 55 \AA~to 35 \AA~in steps of 2 \AA~and
  from 35 \AA~to 24 \AA~in steps of 1 \AA. While decreasing the pair separation,
  at each pair separation, equilibration run of 0.5 ns was done. After obtaining
  initial system at each pair separation, independent equilibration and
  production runs of 4 ns and 5 ns, respectively, were performed. Following the same protocol mentioned above, PMFs for \m~and \a~functionalized 
  nanoparticles in vacuum and SPC/E water are also obtained at 275 K and 1 atm in order to understand the effect of temperature.

\section{Results}
\label{sec:result6}
 \subsection{Bulk Solvent: Structure, Thermodynamics and Transport Properties}
\begin{figure}[h!]
\begin{center}
  \includegraphics[clip=true,width=8.5cm]{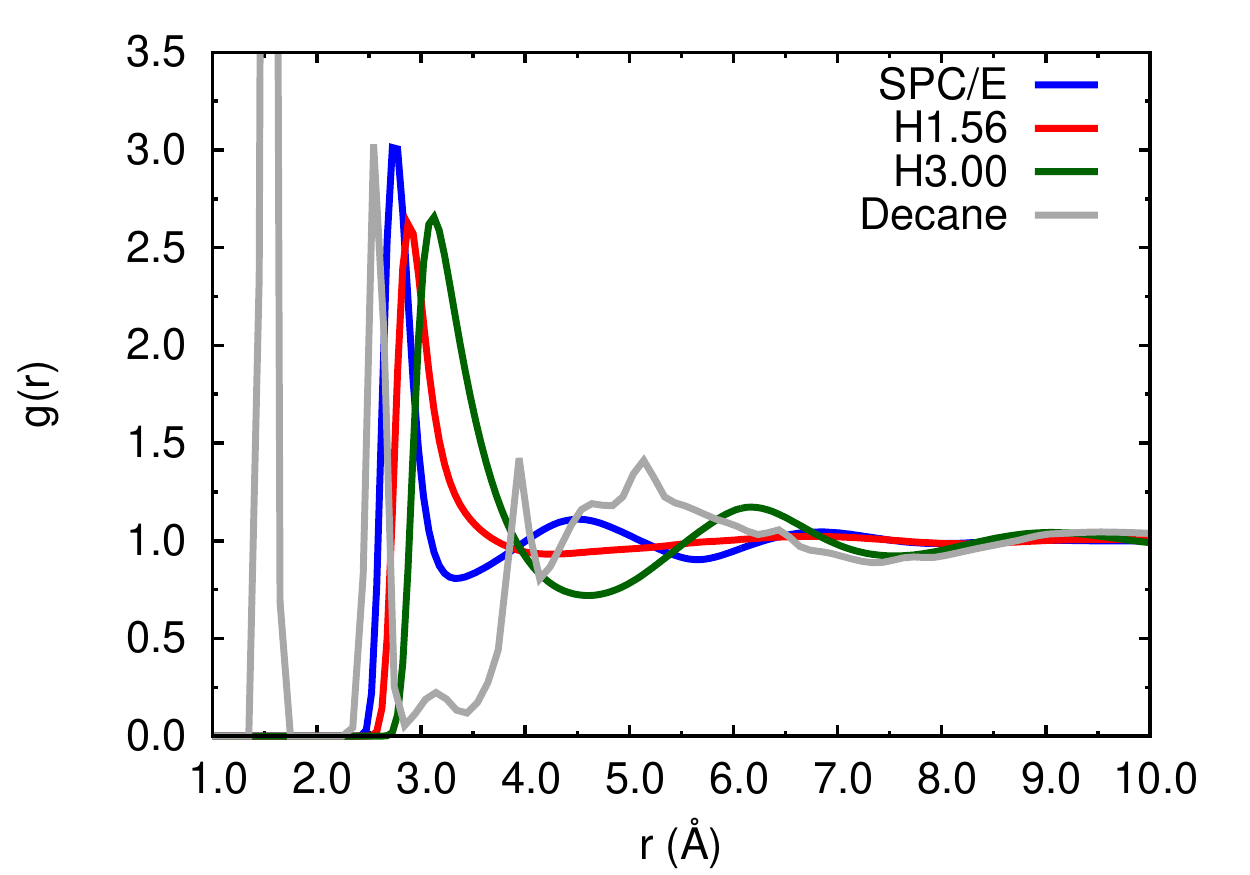}
    \end{center}
    \caption{Comparison of the oxygen-oxygen radial distribution functions
      (RDFs) for modified hybrid water models with $\lambda$ = 1.00, 1.56
  and 3.00 at 300 K and 1 atm. The RDF of decane monomer units including bond, angle and dihedral
  peaks has also been shown at the same temperature and pressure.}
        \label{fig:gss}
\end{figure}

As we have already mentioned, the structure, thermodynamics and transport
properties of solvent play a significant role in the self-assembly process. Hence,
it is very important to understand the solvent properties first.
Figure~\ref{fig:gss} shows the radial distribution function (RDF) of oxygen atoms of
water molecules for modified hybrid water models with $\lambda$ = 1.00, 1.56 and
3.00 at 300 K and 1 atm, where $\lambda$ = 1.00 corresponds to pure SPC/E water. The same figure
also shows the RDF of decane monomer units at the same temperature and pressure, including bond, angle and dihedral
peaks.
The RDFs of oxygen atoms clearly show the change in position and height of the 
first and second peaks with increasing contribution of LJ term relative to the electrostatic term. The position of the first peak for SPC/E water is at
$\approx$ 2.75 \AA, which is clearly less than the van der Waals size parameter of an oxygen
atom, due to hydrogen bonding. The second peak lies at $\approx$ 4.60
\AA, about 1.7 times the distance of the first peak position. This structure is
a typical signature of tetrahedral liquids. As the hybrid potential parameter,
$\lambda$, is progressively increased, the peak positions change and at
$\lambda$ = 3.00, the
first and second peaks are at $\approx$ 3.20 \AA~and $\approx$ 6.40 \AA~, respectively. The second coordination
shell is at twice the distance of the first shell and is typical of a
Lennard-Jones liquid. The non-monotonic change in the first peak height is also
notable with the change in the $\lambda$. At intermediate $\lambda$, there is almost no structure
beyond the first shell. Hence, with an increase in $\lambda$, the liquid structure
shifts from tetrahedral water-like to LJ-like encompassing a liquid with least structure at
intermediate $\lambda$. This can be further understood by calculating various
other thermodynamic and dynamic properties as reported in Table~\ref{tab:solvent}.

The number densities, $\rho_{liq}^*$, reported in Table~\ref{tab:solvent} for different solvents
are almost similar ($\approx$ 0.03 \AA$^{-3}$). The number densities for
modified hybrid water models have been calculated by taking the number of oxygen atoms, whereas 
in case of decane, each decane monomer unit is taken as a
single entity. The correlation
coefficient ($R = {\left<\Delta W\Delta U\right>}/{\sqrt{\left<\left(\Delta
          W\right)^{2}\right>} \sqrt{\left<\left(\Delta U\right)^{2}\right>}}$)
          and the slope of the correlation plot ($\gamma_{UW} = {\left<\Delta
          W\Delta U\right>}/{\left<\left(\Delta U\right)^{2} \right>}$) are also
          reported\cite{bpgsd08,bpgs08,sbpgd09,gspbd09,pbsd08,bigpds12,jcd13,bbvsd13}. The correlation coefficient defines the correlation between
          configurational energy and virial for liquids and can be used to judge
          the extent of simple liquid behaviour. Simple liquids show very strong
          configurational
          energy-virial correlations with $R\geq$ 0.90\cite{bpgsd08}. As the $\lambda$
                is increased from 1.00 to 3.00, the $R$ value
                increases from 0.063 to 0.317, indicating a progressive increase
                in simple liquid behaviour. The slope of the correlation plot
                also increases from 0.349 for $\lambda$ = 1.00 to 2.768 for
                $\lambda$ = 3.00. In case of decane, for calculating the
                correlation coefficient, only non-bonded contributions to
                potential energy and virial are considered. The $R$ value is very
                high for decane and this value as well as the corresponding $\gamma_{UW}$ value
                for decane are comparable to simple liquids. High $R$ values also
                indicate that the structure and dynamics of the liquid will be
                dominated by pair correlations. The isothermal
                compressibilities, pair entropies and diffusivities of different solvents have also been
                reported in Table~\ref{tab:solvent}. The isothermal
                compressibilities ($\kappa_\tau$) are calculated using fluctuation formula and
                give the extent to which a liquid can be compressed under
                certain applied pressure. The
                pair entropy can be related to the local structure of the
                liquid and for an atomic mixture with $X_\alpha$ mole fraction
                of species $\alpha$, the pair entropy, $S_2$, is given by $S_2/Nk_B = \sum_{\alpha ,\beta }
                              X_{\alpha }X_{\beta } S_{\alpha\beta}$,
\begin{equation}
                           S_{\alpha\beta} =
                           -2\pi\rho\int_0^\infty\{g_{\alpha\beta}(r)\ln
                                    g_{\alpha\beta}(r) -\left[
                                    g_{\alpha\beta}(r)-1\right] \} r^2 dr 
\end{equation}
  where $g_{\alpha\beta}(r)$ is the pair correlation function (PCF) between
  atoms of type $\alpha$ and $\beta$\cite{jah90,lh92}. The diffusivities were
  calculated using the Einstein relation
  ($D_\alpha=\lim_{t\rightarrow\infty}{\langle|\delta
  {\bf{r}}_i\left(t\right)|^2\rangle}/{6t}$)\cite{hm06}.
  Decane has the highest compressibility with lowest entropy and diffusivity
  among all the solvents used at 300 K and 1 atm, whereas if we consider
  modified hybrid water
                models, the compressibility, pair entropy and diffusivity follow 
                non-monotonic behaviour with the increase in hybrid potential
                parameter, $\lambda$. The non-monotonic changes
                observed in different thermodynamic and dynamic properties of
                modified hybrid water models are in sync with the changes
                observed in the structure defined by radial distribution functions
                as shown in Figure~\ref{fig:gss}. The liquid with intermediate
                $\lambda$ value (H1.56), has the least structure beyond the first shell and is evident from
                high compressibility, pair entropy and diffusivity values as compared to other
                modified hybrid water models. In our earlier paper on modified hybrid water models, we have 
		reported the change in LJ and electrostatic contributions to the configurational energy with the change in $\lambda$\cite{pc14}.
		With the increase in $\lambda$, the LJ contribution to configurational energy becomes more negative 
		and the electrostatic contribution becomes more positive\cite{pc14}. For H3.00, the electrostatic contribution 
		to the configurational energy is dominant (but more positive as compared to SPC/E and H1.56) and the hydrogen-bond energy 
		is estimated to be 70$\%$ of the value for SPC/E water. Therefore, the system can be treated as a moderately polar one.
\begin{table}[h!]
\caption{The structure, thermodynamic and dynamic properties of all the solvents
  used in this study at 300 K and 1 atm. The number density ($\rho^*_{liq}$),
  correlation coefficient ($R$), slope of the correlation plot ($\gamma_{UW}$),
isothermal compressibility ($\kappa_\tau$), pair entropy (S$_2$/Nk$_B$) and
diffusivity (oxygen of water/decane monomer unit of different solvents) are tabulated.}
\label{tab:solvent}
\centering
\setlength{\tabcolsep}{0.20cm}
\renewcommand{\arraystretch}{1.1}
\begin{tabular}{l c c c c c c}
\hline
\hline
Solvent & $\rho^*_{liq}$ (\AA$^{-3}$) & R & $\gamma_{UW}$ &
$\kappa_\tau$ (10$^{-10}$ Pa$^{-1}$)& S$_2$/Nk$_B$ & Diff. (10$^{-9}$
m$^2$ s$^{-1}$)  \\
\hline
SPC/E & 0.033 & 0.063 & 0.349 & 4.521 & -2.175 & 2.853 \\
H1.56 & 0.032 & 0.152 & 0.935 & 4.677 & -2.085 & 5.835 \\
H3.00 & 0.032 & 0.317 & 1.991 & 2.768 & -2.558 & 5.512 \\
Decane & 0.031 & 0.911 & 5.421 & 13.584 & -3.836 & 2.275 \\
\hline
\hline
\end{tabular}
\end{table} 
\subsection{Solvation of a Single Nanoparticle}
In this section, we study the solvation of three types of end group functionalized
nanoparticles in various solvent medium and infer the effects caused due to the reorganization of 
solvent and ligand in presence of each other. 
\subsubsection{Effect of Solvent Reorganization}
\begin{figure}[h!]
\begin{center}
  \includegraphics[clip=true,trim=6.1cm 9.7cm 6.1cm 8.2cm,width=8.5cm]{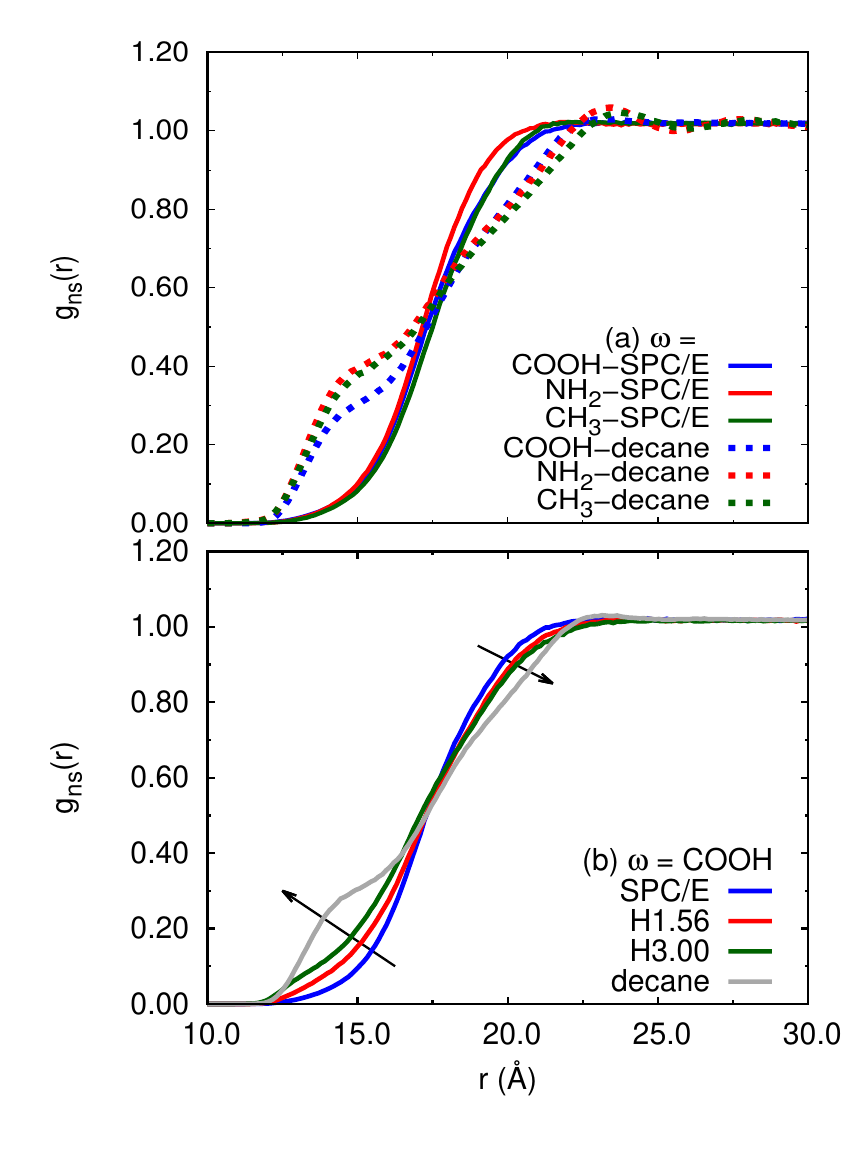}
    \end{center}
    \caption{Pair correlation function, $g_{ns}(r)$: (a) for different
  end groups functionalized nanoparticles in SPC/E water and decane, (b) for \auc~nanoparticle 
  in different solvents at 300 K and 1 atm.}
        \label{fig:gns}
\end{figure}

The reorganization of water molecules and decane monomers around the \aum, \aua~and \auc~nanoparticles can be described by the pair correlation function (PCF),
$g_{ns}(r)$. Figure~\ref{fig:gns}(a) shows the pair
correlation function for all three types of nanoparticles solvated in water and decane at
300 K and 1 atm. 
The end group functionalization of nanoparticles does not seem
to have a substantial effect on the arrangement of water molecules, since the PCFs 
in SPC/E water, show little variation with functionalization, but the solvation
properties related to integral of $g_{ns}(r)$ are expected to show significant
differences. The $g_{ns}(r) <$ 1.0, even in the soft
corona region of the nanoparticles. This is due to the greater solvent-solvent
interactions compared to solute-solvent interactions. This type of PCF has already been
observed in the literature for solvation of \aum~nanoparticle in decane and other
organic solvents at very high densities, whereas the solvation of \aum~nanoparticle in ethane and propane at lower densities close to critical isotherm
show well defined solvation shell ($g_{ns}(r) >$ 1.0)\cite{ysac16,yac17}. Decane solvent shows
better penetrability at lower $r$ values compared to water due to the 
favorable interactions between methylene units of
ligand and decane monomer units.
In decane, \aum~and \aua~nanoparticles show similar and greater solvent
penetration at lower $r$ as compared to \auc~nanoparticle.
To understand the penetration effect of
solvent, we have also reported the PCFs for \auc~nanoparticle in modified hybrid water solvents where the
repulsion-dispersion interaction is
enhanced as compared to the electrostatic interaction as shown in Figure~\ref{fig:gns}(b). With the increase in
$\lambda$, the repulsion-dispersion contribution to the total potential energy of
the solvent increases. The penetrability of the solvent at lower $r$ increases with increase in $\lambda$ and
is highest in case of decane, where the solute-solvent interaction is even more
favorable compared to H3.00 modified hybrid water model. 
Similar effects with
the change in the solvent can also be seen for \aum~and \aua~nanoparticles (data not shown here).

\begin{figure}[h!] 
\begin{center}
  \includegraphics[clip=true,trim=4.4cm 12.6cm 4.0cm 10.6cm,width=16cm]{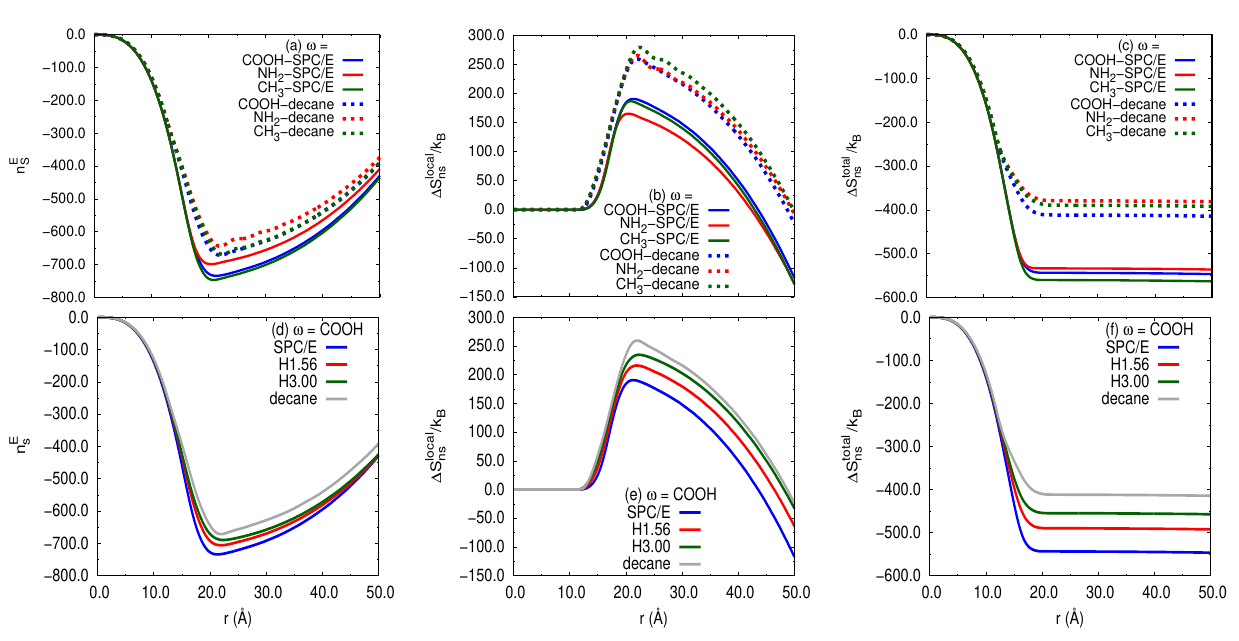}
    \end{center}
    \caption{The running integral of different solvation properties: (a) solvent excess number, $n^E_s$, (b) solvation pair entropy due to
  local ordering of solvents, ${\Delta}S^{local}_{ns}/k_B$ and (d) total entropy of solvation with local and
	long-range correction terms, ${\Delta}S^{total}_{ns}/k_B$, in water and decane at 300 K and 1 atm. (d) $n^E_s$, (e) ${\Delta}S^{local}_{ns}/k_B$ and (f) ${\Delta}S^{total}_{ns}/k_B$ 
	for \auc~nanoparticle in different solvents at 300 K and 1 atm.} 
        \label{fig:gns-quan}
\end{figure} 
The reported pair correlation functions, $g_{ns}(r)$, do not approach unity even when $r \,\to\, \infty$ due to use
of finite system size and NVT conditions, which results in a significant deviation of
$1/N$ from unity. Hence, all the solvation quantities, which depend on the
Kirkwood-Buff integral do not converge, however, the relative values can provide
significant information. 

Figure~\ref{fig:gns-quan} shows the running integral of the solvation properties for all three types 
of nanoparticles in water and decane and for \auc~nanoparticle in different solvents at 300 K and 1 atm. It has already been shown
in literature that the $n^E_s$ is positive and shows anti-correlating behavior
w.r.t local entropy (${\Delta}S^{local}_{ns}/k_B$) for solvation of alkanethiol
passivated gold nanoparticle in ethane and propane at low and intermediate densities near the
critical isotherm, where well defined solvation shell (g$_{ns}$(r) $>$ 1.0) is
observed\cite{nyjc12,yac17}. At very
high densities, no well defined solvation shell is
observed (g$_{ns}$(r) $<$ 1.0), the solvent excess shows negative values and the anti-correlation does
not seem to hold true in that case\cite{nyjc12,ysac16,yac17}, which is similar to our study.
In our study, $n^E_s$ is
maximum (less negative) in decane for all the nanoparticles i.e. the affinity of
nanoparticles for decane solvent is highest among all the solvents used in this
study, which is expected, since the
penetration of decane into the ligand shell is maximum.
With the increase in $\lambda$ or repulsion-dispersion interaction, the solvent
penetration increases inside the ligand soft corona as compared to H1.00 as seen in Figure~\ref{fig:gns}(b), which
gives less negative $n^E_s$ value in H3.00 than H1.56 and H1.00 i.e., the nanoparticles are
better solvated in H3.00 as compared to H1.56 and H1.00. In all the solvents, the $n^E_s$
value is less negative for
\aua~than \aum~and \auc~nanoparticles. Temperature also affects the n$^E_s$ value. At lower
temperature, the affinity of solute for any solvent is more than at higher
temperature (data not shown here). 

We have also estimated the local
(${\Delta}S^{local}_{ns}/k_B$) and total entropy
(${\Delta}S^{total}_{ns}/k_B$) as shown in Figures~\ref{fig:gns-quan}(b,e),~\ref{fig:gns-quan}(c,f), respectively. The local entropy, as
well as the total entropy, is generally maximum in decane even when the solvent penetration
inside the ligand shell is maximum for decane. In
modified hybrid water models, the local entropy and the total entropy increase
with the increase in $\lambda$. A significant part of the variation in
total entropy arises due to local entropic contributions.

\begin{figure}[h!]
\begin{center}
  \includegraphics[clip=true,width=8.5cm]{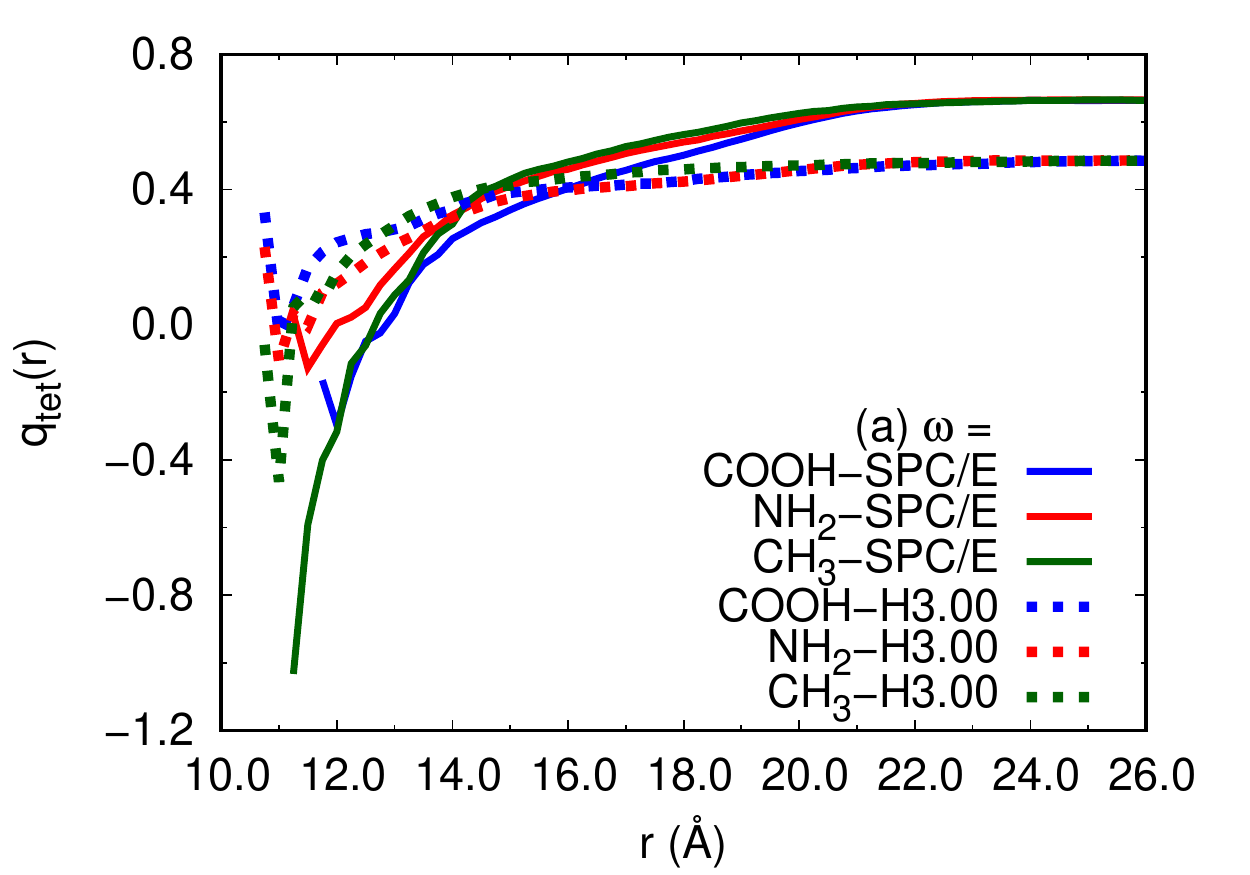}
    \end{center}
    \caption{Comparison of the average local tetrahedral order of waters,
      $q_{tet}(r)$, as a
    function of the distance from the center of mass of the $Au_{140}$ nano-core in
  H1.00 or SPC/E and H3.00 at 275 K and 1 atm. Solid and dashed lines are used
for nanoparticles solvated in H1.00 and H3.00 modified hybrid water model, respectively.}
        \label{fig:qtetr}
\end{figure}

In order to understand the behaviour of water molecules in the presence of
end group functionalized nanoparticles, we attempt to quantify the changes in the 
local ordering of water molecules by calculating the local tetrahedral order
metric, $q_{tet}$. The
breakdown of the tetrahedral, hydrogen bond network of water in the presence of
different solutes can be best monitored using
the local tetrahedral order parameter\cite{ch98,ed01,nc13,ht12}. 
The local tetrahedral order parameter, {$q_{tet}$}, associated with an atom
$i$, is
 defined as \cite{ch98,ed01} 
 \begin{equation}
   q_{tet,i} = 1 -\frac{3}{8}\sum_{j=1}^3\sum_{k=j+1}^4 \left(\cos \psi_{jik}+
   \frac{1}{3}\right)^2  
     \label{eqn:qtet}
   \end{equation}
   where $j,k$ are the nearest neighbors to the central atom, $i$ of a given
   molecule. $\psi_{jik}$ is the angle formed between the bond vectors ${\bf
   r}_{ij}$ and ${\bf
   r}_{ik}$. For perfect tetrahedrality, the value of $q_{tet}$ is 1. While
   calculating $q_{tet}$, we have considered oxygen atom of water molecules and no other
   atoms of the passivated nanoparticle. Figure~\ref{fig:qtetr} shows
   the local tetrahedral order of waters, $q_{tet}(r)$, as a function
   of the 
   distance from the center of mass of the gold core in SPC/E or H1.00 and H3.00
   at 275 K and 1 atm, since the effect is more prominent at low temperatures. 
   
\begin{figure}[h!]
\begin{center}
  \includegraphics[clip=true,trim=6.1cm 10.2cm 6.2cm 8.3cm,width=8.5cm]{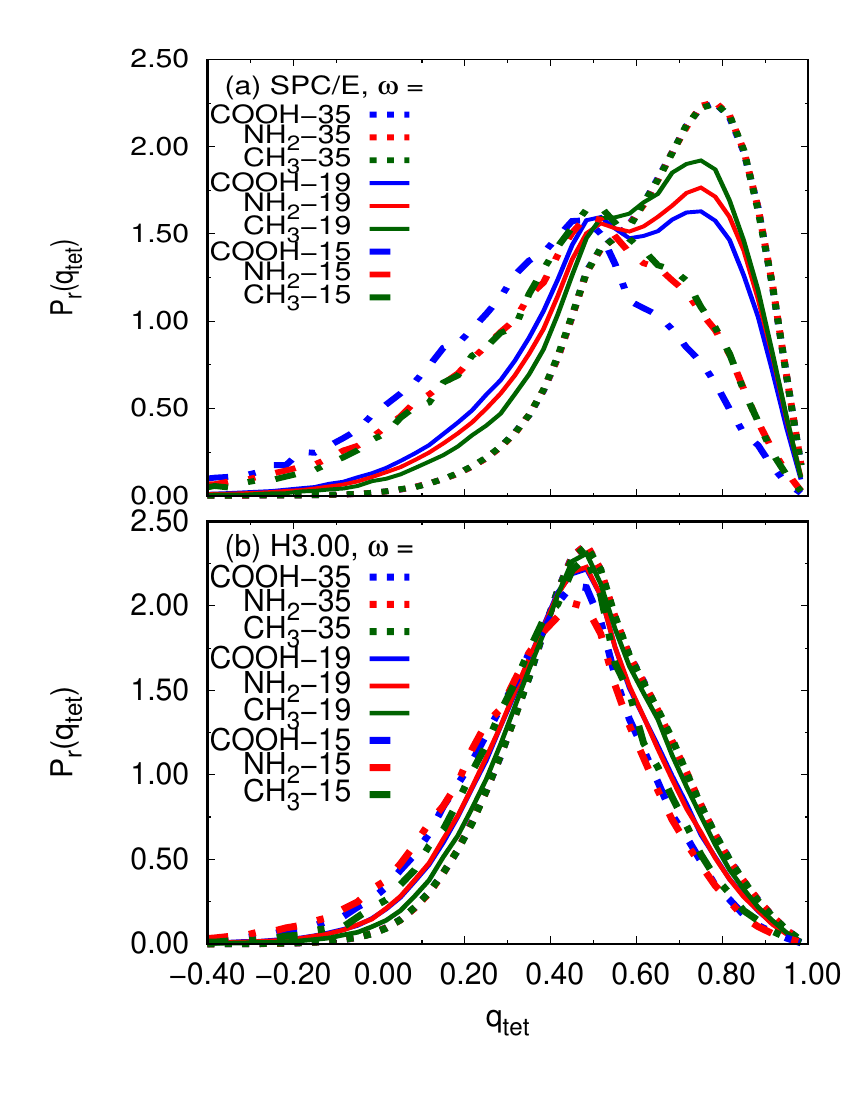}
    \end{center}
    \caption{Conditional tetrahedral distributions, $P_r(q_{tet})$, calculated for water molecules lying within
        distances of 35-36 \AA, 19-20 \AA, and 15-16 \AA~ from the center of mass of
      the gold core for (a) SPC/E and (b) H3.00 water at 275 K and 1 atm.}
        \label{fig:pqtet}
\end{figure}

   At very large distances ($r >$ 24 \AA) from the center of mass of the
   gold core, the effects due to the presence of nanoparticles as a solute are
   insignificant and the tetrahedral order corresponds to the bulk value for
   SPC/E and H3.00 i.e., 0.664 and 0.485, respectively, at 275 K and 1 atm. The
   bulk value for SPC/E is high compared to H3.00 because of the differences in
   the hydrogen bond strengths of the two liquids. $q_{tet}(r)$ decreases only when $r$ is 
   close to the periphery of the passivated
   nanoparticle. The decrease in $q_{tet}$ with $r$ is more pronounced for SPC/E water
   compared to H3.00. At low $r$, the decrease is more prominent, probably because of
   the less number of SPC/E water compared to H3.00 water inside the
   ligand shell as evident from the degree of penetration of solvents as shown in
   Figure~\ref{fig:gns}(b). The \auc~nanoparticle affects
   the tetrahedral ordered network of water to a greater extent as compared to
   other end group functionalized nanoparticles used in this study as apparent from the $q_{tet}(r)$
   data for all other nanoparticles in SPC/E water. This can be well understood by plotting the
   conditional tetrahedral distribution, $P_r(q_{tet})$, plots.
   Figure~\ref{fig:pqtet} shows the conditional
   tetrahedral distributions, calculated for water molecules lying within
   distances of 35-36 \AA, 19-20 \AA, and 15-16 \AA ~from the center of mass of the
   gold core for SPC/E and H3.00 water at 275 K and 1 atm. For SPC/E water, the
     $P_r(q_{tet})$ plots for all the three end group passivated nanoparticle
     system at 35-36 \AA~distance overlap with each other and show a high
   tetrahedrality peak at $\approx$ 0.80 and a very small bump at slightly low
   tetrahedrality. Thus, there is
   no effect due to nanoparticles or the end group functionalization on the
   tetrahedral structure of water molecules and the $q_{tet}(r)$ resembles the
   tetrahedral order of bulk water. At
   19-20 \AA ~distance, the water molecules are near or just inside the ligand
   soft corona and hence the effect due to end group functionalization can be
   clearly seen. The degree of interaction of the functionalized group with water molecules follows the
   order COOH $>$ NH$_2$ $>$ CH$_3$ and hence the degree of breaking the tetrahedral
   order of water also follows the same order as noticeable in the $P_r(q_{tet})$
   plots at intermediate $r$. At this distance, a clear bimodal distribution can
   be seen. In case of H3.00 modified hybrid water as the solvent, the
   tetrahedral structure of the liquid is already broken due to low hydrogen
   bond strength and hence all the $P_r(q_{tet})$ plots look alike.
\subsubsection{Effect of Ligand Reorganization}  
\begin{figure}[h!] 
\begin{center}
  \includegraphics[clip=true,trim=6.1cm 9.7cm 6.2cm 8.1cm,width=8.5cm]{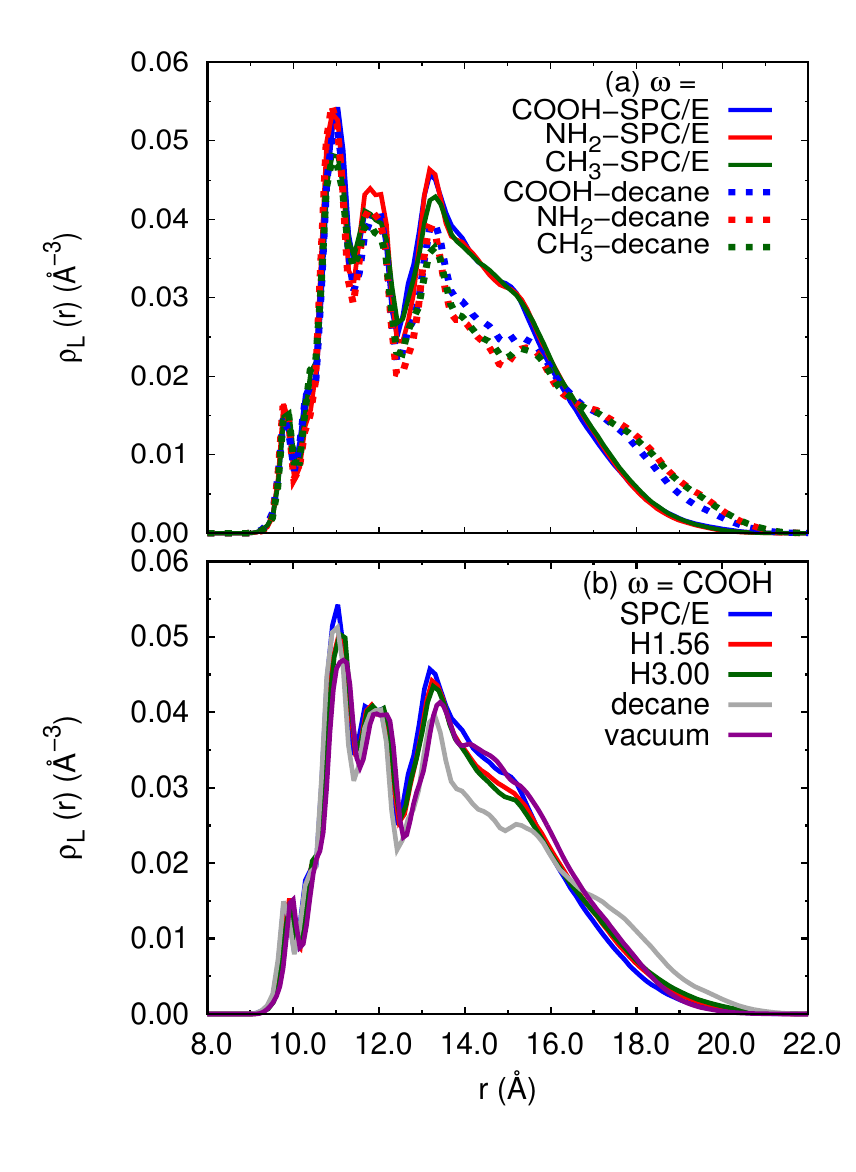}
    \end{center}
    \caption{Radial density profiles, $\rho_L(r)$, of ligand chain monomers (for
      nine CH$_2$ groups only, which are common in all the end group
      functionalized nanoparticles) in (a) water and decane and (b) solvents of
    varying degrees of repulsion-dispersion and electrostatic interactions for
  \auc~nanoparticle at 300 K and 1 atm.}
        \label{fig:gnn}
\end{figure}

In this section, we attempt to quantify the structure of the ligand shell of
the three end group functionalized nanoparticles used in
this study in the presence of liquids with varying degrees of
repulsion-dispersion and electrostatic interactions. The ligands are free to move in the presence of solvent and hence,
contribute to the total free energy of solvation. The shape and size of the
passivated nanoparticle also affects the self-assembly process.
Figure~\ref{fig:gnn}(a) shows the structure of the ligand shells, characterized
by radial density profiles (RDP), $\rho_L(r)$, in SPC/E water and decane at 300 K and 1 atm. 
The nine 
methylene groups, which
are common in all type of ligands are used to compute the radial density
profiles with respect to the center of mass of
the gold core. In SPC/E water, the structures of all the different types of
ligand shells are almost similar, except slight differences in peak heights for
\m~ligand shell as compared to \a~and \c~ligand shells. The similarity
of RDPs in SPC/E water may be attributed to very less penetration into the ligand soft
corona by water molecules. The ligand structure is more open and stretched in
decane compared to SPC/E, probably due to more penetration of decane solvent
inside the ligand shell as compared to SPC/E. The methylene units are stretched up
to $\approx$ 21
\AA~ and $\approx$ 20 \AA~in decane and SPC/E, respectively. The differences between the
structure of different ligand shells are also more prominent in decane than
SPC/E. Figure~\ref{fig:gnn}(b) shows the \c~ligand shell structure
in different solvents and vacuum. As we have already mentioned, with the increase in interaction between solute
methylene units and solvent, the penetration of the solvent inside the ligand
shell increases. This increase affects the ligand shell structure and the
ligands try to stretch more, which increases the probability of methylene units
at $r >$ 16 \AA~ and decreases the probability of methylene units at intermediate $r$
values from the center of mass of the gold core.

\begin{figure}[h!]
\begin{center}
  \includegraphics[clip=true,trim=6.1cm 10.0cm 6.2cm 8.4cm,width=8.5cm]{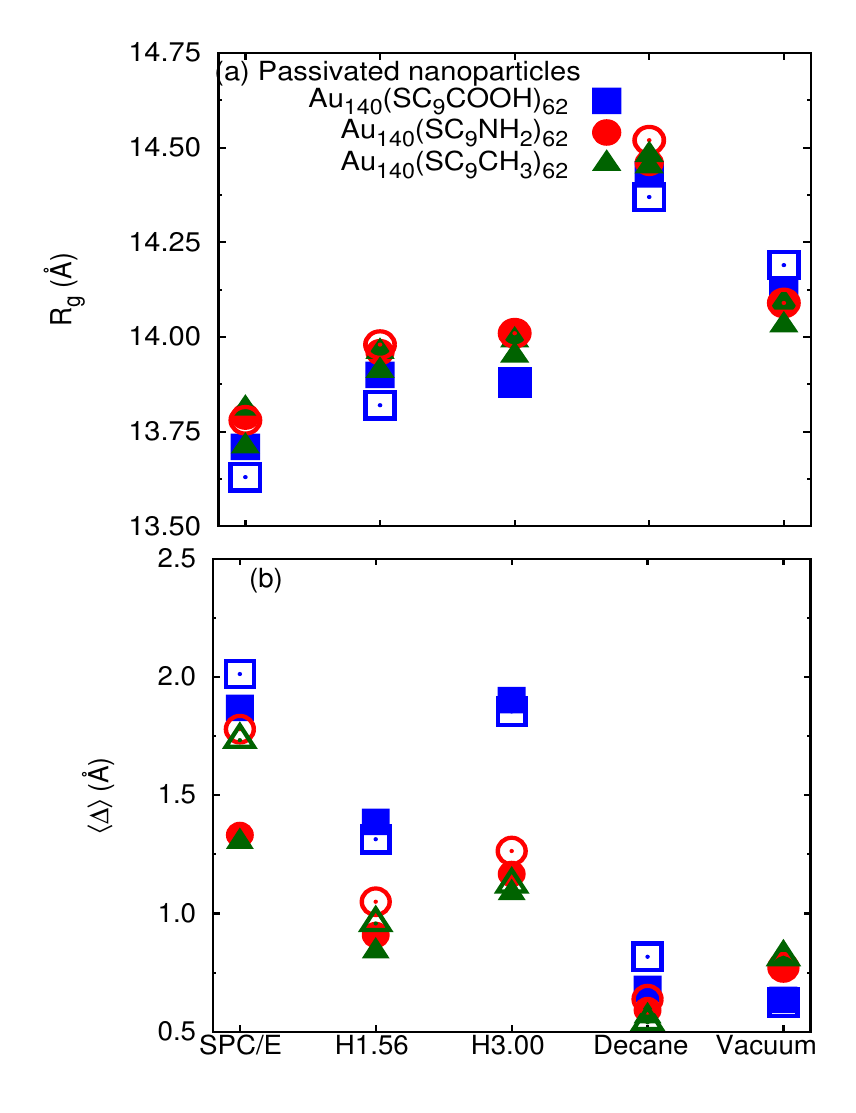}
    \end{center}
    \caption{(a) Radius of gyration between the center of mass of the gold nano-core
    and the SH, CH$_2$ groups of the passivating ligands and (b) $\langle \Delta
    \rangle$,
  where, $\Delta$, denotes the magnitude of the vector connecting the center of
mass of the gold nano-core and the SH, CH$_2$ groups of the ligand shell. The
solid and open points are used for data at 300 and 275 K at 1 atm, respectively.}
        \label{fig:rgr}
 \end{figure}

In order to quantify the structure of the ligand shell further, we have calculated
the radius of gyration ($R_g$), which provides a quantitative measure of the expanded state
of the ligand shell and we attempt to relate the $R_g$ data to the RDPs and the
inherent anisotropy of the passivated nanoparticles. $R_g$ has been
calculated using the average distances of sulfur and methylene groups of all the
ligand chains from the center of mass of the gold core and are reported in
Figure~\ref{fig:rgr}(a). Since all three types of ligand shells are most stretched in
decane solvent, as apparent from Figure~\ref{fig:gnn}, the highest $R_g$ value
for all three types of ligands is seen in decane solvent. The $R_g$ values
follow the order, decane $>$ vacuum $>$ H3.00 $\approx$ H1.56 $>$ SPC/E. The $R_g$
value for \c~ligand shell is in general slightly less as compared to
\a~and \m~ligand shells except in vacuum, probably because the
COOH group is a bulky group compared to CH$_3$, NH$_2$ groups and under the effect of this bulky group and pressure of the
solvent, the ligand
chains fold slightly. Temperature also significantly alters the $R_g$ values, but this
effect cannot be generalized.

\begin{figure}[h!]
\begin{center}
  \includegraphics[clip=true,trim=0cm 16.0cm 4.1cm 4.5cm,width=16cm]{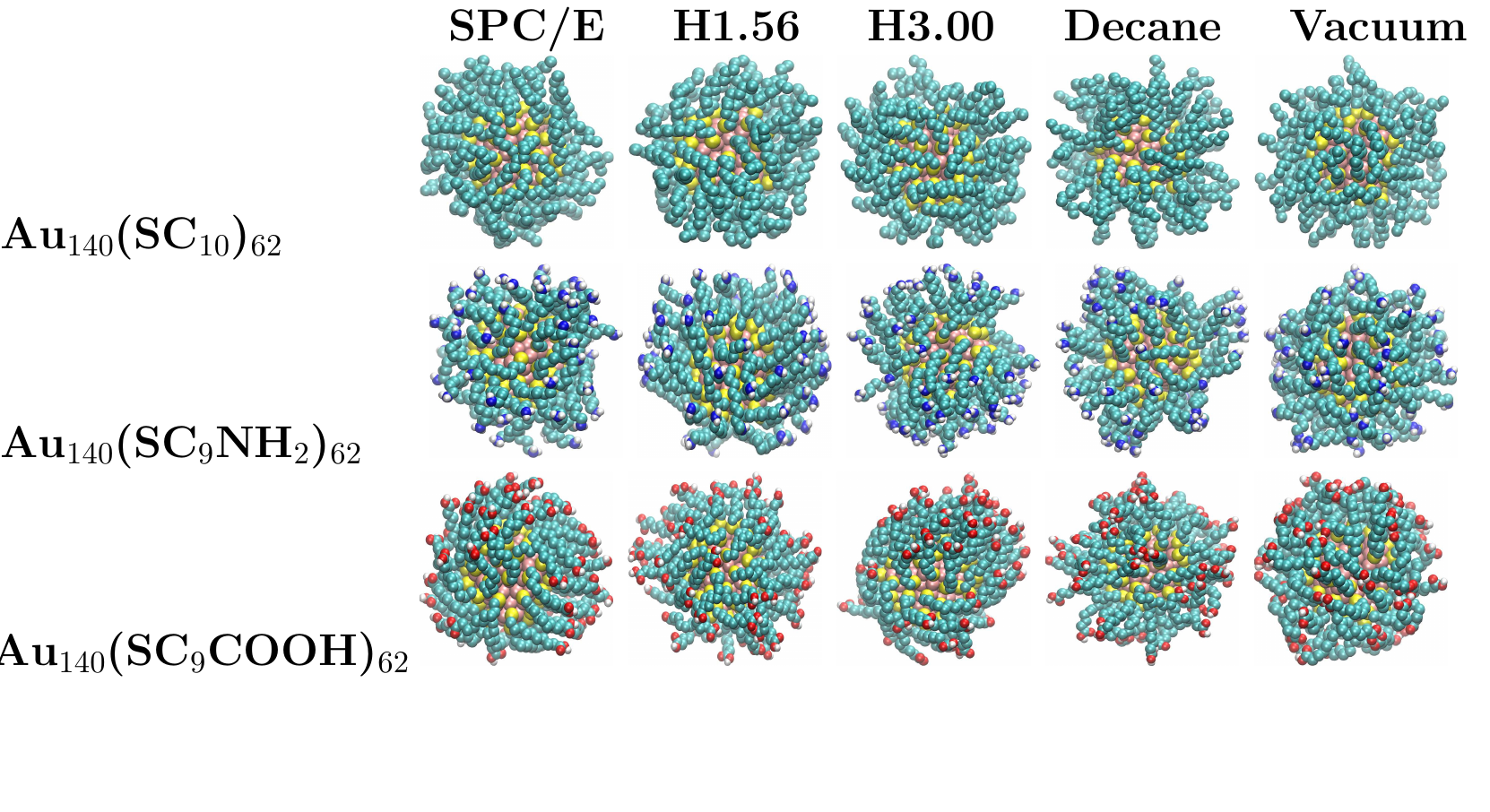}
    \end{center}
    \caption{Snapshots of different end group functionalized nanoparticles in
    solvents of varying interactions at 300 K and 1 atm. Solvent molecules are not
  shown for clarity. Sulfur atoms are shown in yellow, carbon in cyan, oxygen in
red, nitrogen in blue and hydrogen in white.}
        \label{fig:snap1}
\end{figure}

Figure~\ref{fig:rgr}(b) shows the $\langle \Delta \rangle$, where $\Delta$
denotes the magnitude of the vector connecting the center of mass of the gold
core and the sulfur and methylene groups of the ligand shell. The sulfur and
methylene groups are considered to effectively compare all the three types of
ligand shells. The $\Delta$ value represents the degree of inherent anisotropy
present in the passivated
nanoparticles, more the $\Delta$ value, more is the anisotropy. The highest
anisotropy is seen in SPC/E and H3.00 water and least in decane for all the
three types of
ligand shells. The anisotropy follows the order, SPC/E $\approx$ H3.00 $>$ H1.56
$>$ vacuum $\approx$ decane, which is almost inverse of the order observed for $R_g$. The 
anisotropy of the ligand shells depend on the competition between different types of interactions such as
ligand-solvent interaction, ligand-ligand interaction, the free volume available to
each chain and slightly on the mobility of the solvent. In H1.56, the ligand
shells show low anisotropy compared to SPC/E may be due to greater ligand-solvent interaction
and also due to high diffusivity of H1.56 solvent as given in Table~\ref{tab:solvent}.
The ligand shells in H3.00 show higher anisotropy compared to H1.56 irrespective
of better ligand-solvent interaction as evident from greater solvent
penetration and higher solvent excess for H3.00, possibly due to
low diffusivity of solvent compared to H1.56. The anisotropy of 
\c~ligand shell is generally
high in all media studied, except in vacuum, which is also inverse of what we observed in
case of $R_g$ values. The high values of anisotropy in \c~ligand shell in SPC/E
and modified hybrid water models can be
attributed to the formation of ligand-ligand or ligand-solvent hydrogen bonds. At low temperature,
generally, the anisotropy increases except in certain cases reported in
this study. For a better understanding of the extent of anisotropy, we have also shown the probability distribution of 
$\Delta$ in Figure~S4 of the supplementary material\cite{supple-nano}. 
All the quantities calculated above to quantify the effect of ligand reorganization have taken into account the part of ligands which 
are common in all ligands, hence even a small change is of importance and will magnify the difference if the entire ligand is taken into account.
Figure~\ref{fig:snap1} shows the snapshots of all three types of
passivated nanoparticles solvated in different solvent media. The snapshots
clearly show the significant qualitative differences in the structure of the
coatings or coating asymmetry.

In Table~\ref{tab:schlitter}, we have reported the upper bound to the ligand
shell configurational entropy using Schlitter's method\cite{js93}. This method was
originally developed to calculate the configurational entropy of biomolecules. The soft ligand
shell is expected to reorganize differently in different solvent media and the
entropy associated with the organization of the ligand shell can be computed using
the covariances of fluctuations in positions of ligand atoms. The entropy for all
three nanoparticles is maximum in vacuum and follows the order, vacuum $>$
decane $>$ H1.56 $>$ H3.00 $\approx$ SPC/E. Nanoparticles in decane have greater
ligand shell configurational entropy than all other water models due to greater
penetration of decane solvent into the ligand shell. H3.00 solvent penetrates the ligand shell to a greater extent than
H1.56 solvent, but the ligand shell configurational entropy is more in H1.56 
compared to H3.00. The H1.56 solvent has significantly greater diffusivity than H3.00 and SPC/E as shown in Table~\ref{tab:solvent}. The 
greater diffusivity contributes to the enhanced reorganization of ligand shell and hence
the S$_L$ is more for nanoparticles solvated in H1.56 than H1.00 and H3.00. Therefore, S$_L$ also depends on the local fluctuations of the solvent medium\cite{gcb16,gncb16}. If we
compare the S$_L$ of the three nanoparticles in all the solvents used, we find
that S$_L$ of \aum~ $>$ \aua~ $>$ \auc. The S$_L$ for \auc~is minimum may be due
to the formation of hydrogen bonds between the two COOH groups of different
ligands of the same nanoparticle, which
restricts the ligand motions, when the
nanoparticle is solvated in non-polar liquid like decane. The COOH
group from the \auc~nanoparticle, when solvated in polar liquid like water is expected to form hydrogen bonds with water, which may also
restrict the ligand motions and decreases the ligand shell configurational entropy. 
\begin{table}[h!]
  \caption{Ligand shell configurational entropy, $S_L$ (J K$^{-1}$ mol$^{-1}$), due to the reorganization of the
ligands in the presence of different solvents at 300 K and 1 atm.}
\label{tab:schlitter}
\centering
\setlength{\tabcolsep}{0.20cm}
\renewcommand{\arraystretch}{1.1}
\begin{tabular}{l c c c c c }
\hline
\hline
NP/Solvent & SPC/E & H1.56 & H3.00 & Decane & Vacuum \\
\hline
\aum & 58.98 & 59.59 & 58.79 & 59.90 & 63.95 \\
\aua & 53.81 & 54.73 & 54.59 & 55.56 & 58.54 \\
\auc & 48.50 & 49.88 & 49.73 & 50.37 & 53.82 \\
\hline
\hline
\end{tabular}
\end{table}
 
\subsection{Effective Interactions Between a Pair of Nanoparticles}
In this section, we have computed the isotropic potential of mean force (PMF)
between two identical end group functionalized ligand passivated
nanoparticles in vacuum, SPC/E water and H3.00 modified hybrid water. The
gold nanoparticles self-assemble to form body-centered cubic (bcc), face-centered
cubic (fcc) and hexagonal close-packed (hcp) ordered structures depending on the
ligand and solvent quality, indicating that the
PMF between a pair of nanoparticles is isotropic in nature. However, the gold
nanoparticles depending on the ligand, solvent quality and temperature, also
form amorphous agglomerates, suggesting the presence of some angular dependence
in addition to pair separation distance in overall free energy between a pair of
nanoparticles. So, we have also characterized the fluctuation driven anisotropy
as a function of the distance
between the two nanoparticles in order to understand the importance of
anisotropy or orientational dependence arising due to ligand corona
fluctuations.

\begin{figure}[h!]
\begin{center}
  \includegraphics[clip=true,trim=6.1cm 7.4cm 6.2cm 5.8cm,width=8.5cm]{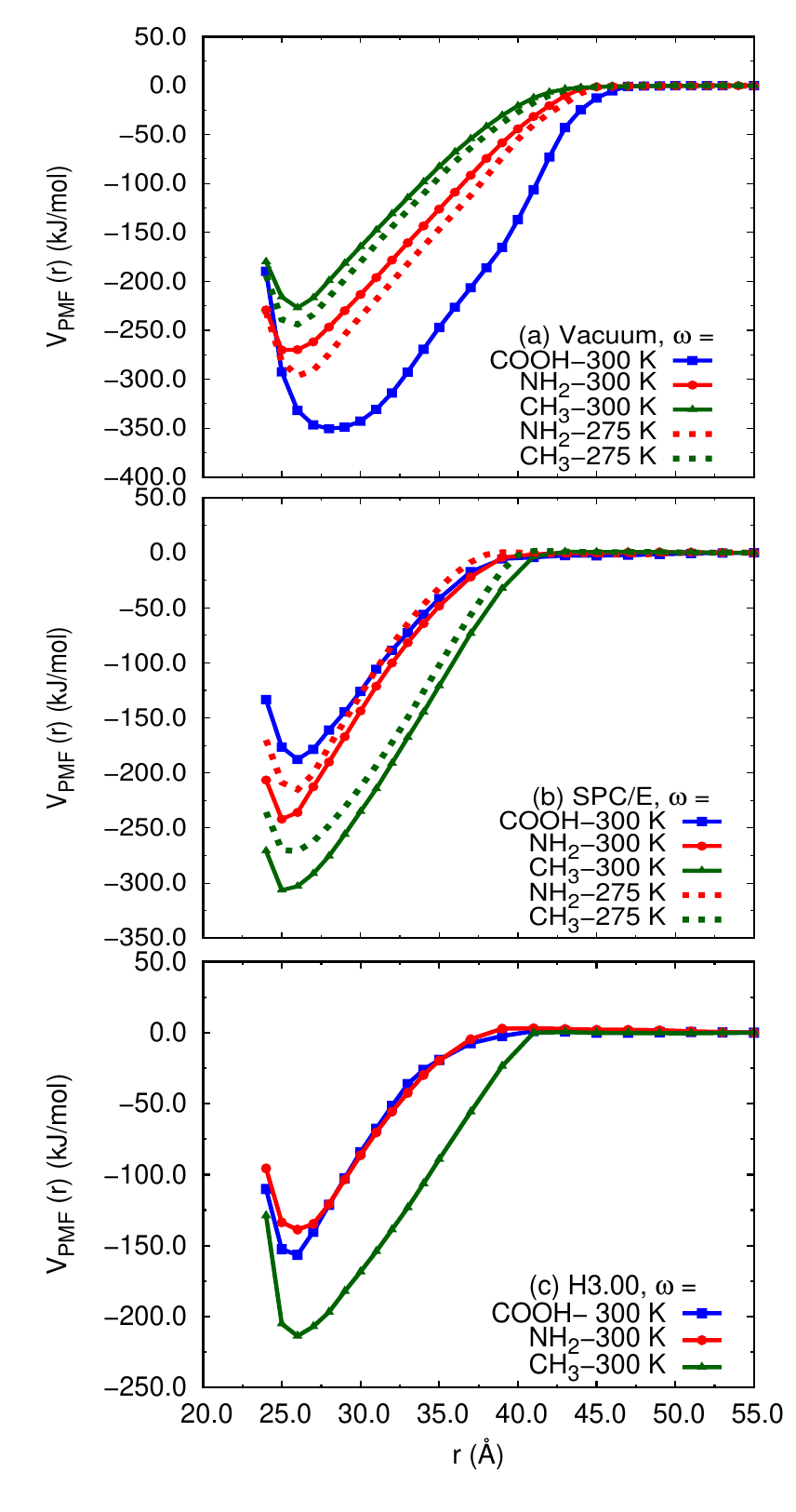}
    \end{center}
    \caption{The isotropic potential of mean force, $V_{PMF}(r)$, as a function of
  pair separation, $r$, between two identical end group functionalized
nanoparticles in (a) vacuum, (b) H1.00 or SPC/E and (c)
H3.00. Solid lines are for 300 K and dashed lines are for 275 K.}
        \label{fig:pmf}
\end{figure} 
Figure~\ref{fig:pmf} shows the isotropic PMF between two identical nanoparticles (\aum~or \aua~or \auc). In all 
three media (vacuum, SPC/E water, H3.00 modified hybrid  
water), the PMF profiles are qualitatively similar and are in contrast
to the PMF profile for \aum~nanoparticle in organic solvents like hexane, octane and
decane as reported in our earlier study\cite{ysac16}. 
All the PMF profiles in vacuum, SPC/E and H3.00 are characterized by a deep attractive well,
whereas the PMF profile of \aum~in organic solvent like decane shows repulsive
PMF profile, which indicates that decane acts as a good solvent condition,
whereas SPC/E and H3.00 at $\approx$ 1.0 g/cm$^3$ density represent poor solvent
conditions as compared to decane. The range of interactions is defined as
the maximum pair separation for which the attractive interactions become
observable and it can be approximated by $2R_c+2L$, where $R_c$ is the radius
of gold core (8 \AA) and $L$ is the distance from the surface of the gold core to
a position in space, where the ligand monomer density is effectively zero. The $L$ value
can also be approximated using the extended length ($L_{ext}$) of an alkylthiol,
$L_{ext} = (n+1) \times 1.2$, where $n$ is the number of carbon atoms in the
alkylthiol. The range of
interactions is slightly high in vacuum than the approximated value of 
$2R_c+2L \approx$ 42.4 \AA~for \aum~nanoparticle, but in
solvent, it fits well with the approximated value. All the PMF profiles can be
fitted to a Morse function,
$D_e\left(\left(1-exp\left(-a\left(r-r_e\right)\right)\right)^2-1\right)$, where
$a = \alpha/r_e$, $\alpha$ is the dimensionless range parameter, $r_e$ is the
equilibrium pair separation and $D_e$ is the
well depth at $r_e$. The fitting  parameters are reported in
Table~\ref{tab:morsefit}. The attractive well indicates that the energetics 
due to ligand-ligand interactions dominate over the entropic effect due
to interdigitation of ligands. In vacuum, the well depth ($D_e$) is maximum and
minimum for \auc~and \aum~nanoparticle, respectively, since in the absence of
solvent, ligand-ligand energetic interaction for \c~ligands is more as compared to
\m~ligands. In SPC/E water, the
variation of well depth follows the reverse order as in vacuum. The well depth for
\aum~nanoparticle is maximum and minimum for \auc~nanoparticle solvated in
SPC/E. This can be explained on the basis of greater COOH-water than CH$_3$-water
interactions. So, the
ligand-ligand energetic  interaction is more for \aum~nanoparticle in SPC/E, which increases 
the attractive well depth as compared
to \auc~nanoparticle. The well depths for all types of nanoparticles
are less in H3.00 solvent due to more solvent penetration into ligand shell
i.e., better ligand-solvent interaction, which decreases the energetic effect and
increases the entropic effect due to ligands. As discussed earlier in case of
solvation of single nanoparticle, the solvent
penetration into ligand shell is maximum for decane i.e. the ligand-solvent interaction is
further enhanced than H3.00 and hence, the PMF profile of \aum~ in decane is
repulsive in nature. The equilibrium
separations $r_e$ are also quite different in vacuum, but the variation in $r_e$
is comparatively less in solvent media as compared to vacuum.
\begin{table}[h!]
  \caption{The summary of fitting parameters for $V_{PMFs}$ at 300 K. The isotropic PMFs
  ($V_{PMFs}$) in vacuum as well as in solvents are
fitted with Morse
function,$D_e\left(\left(1-exp\left(-a\left(r-r_e\right)\right)\right)^2-1\right)$,
where $a = \alpha/r_e$, $r_e$ is the equilibrium pair separation and $D_e$ is the
well depth at $r_e$.}
\label{tab:morsefit}
\centering
\setlength{\tabcolsep}{0.01cm}
\renewcommand{\arraystretch}{1.0}
\begin{tabular}{l c c c c c c c c c}
\hline
\hline
NP/Solvent &
\multicolumn{3}{c}{Vacuum}&\multicolumn{3}{c}{SPC/E}&\multicolumn{3}{c}{H3.00}\\
\cmidrule(l{0.2cm}r{0.2cm}){2-4}
\cmidrule(l{0.2cm}r{0.2cm}){5-7}
\cmidrule(l{0.2cm}r{0.2cm}){8-10}
& D$_e$ & $a$ & r$_e$ &D$_e$ & $a$ & r$_e$ &D$_e$ & $a$ & r$_e$ \\
 & (kJ mol$^{-1}$) & (\AA$^{-1}$) & (\AA) & (kJ mol$^{-1}$) & (\AA$^{-1}$) &
 (\AA) & (kJ mol$^{-1}$) & (\AA$^{-1}$) & (\AA) \\
 \hline
 \hline
\aum & 224.70 & 0.185 & 26.01 & 308.60 & 0.169 & 25.60 & 209.20 & 0.165 & 26.22
\\
\aua & 271.00 & 0.155 & 25.85 & 243.60 & 0.240 & 25.48 & 140.80 & 0.247 & 25.81
\\
\auc & 358.30 & 0.130 & 27.84 & 191.70 & 0.238 & 25.99 & 159.70 & 0.284 &
25.65\\
\hline
\end{tabular}
\end{table}  
To understand the effect of the thermodynamic variable like temperature, we have also reported the isotropic PMFs 
for CH$_3$ and NH$_2$ functionalized nanoparticles in vacuum and water at 275 K in Figures~\ref{fig:pmf}(a) and~\ref{fig:pmf}(b), respectively. 
We find that in case of vacuum, with the decrease in temperature, the attractive well-depth increases for both CH$_3$ and NH$_2$ 
functionalized nanoparticles, whereas in water, the well-depth decreases with the decrease in temperature for both CH$_3$ and NH$_2$ 
functionalized nanoparticles, which is in contrast to vacuum results.

\begin{figure}[h!] 
\begin{center}
  \includegraphics[clip=true,trim=0.5cm 14.6cm 4.7cm 4.3cm,width=16cm]{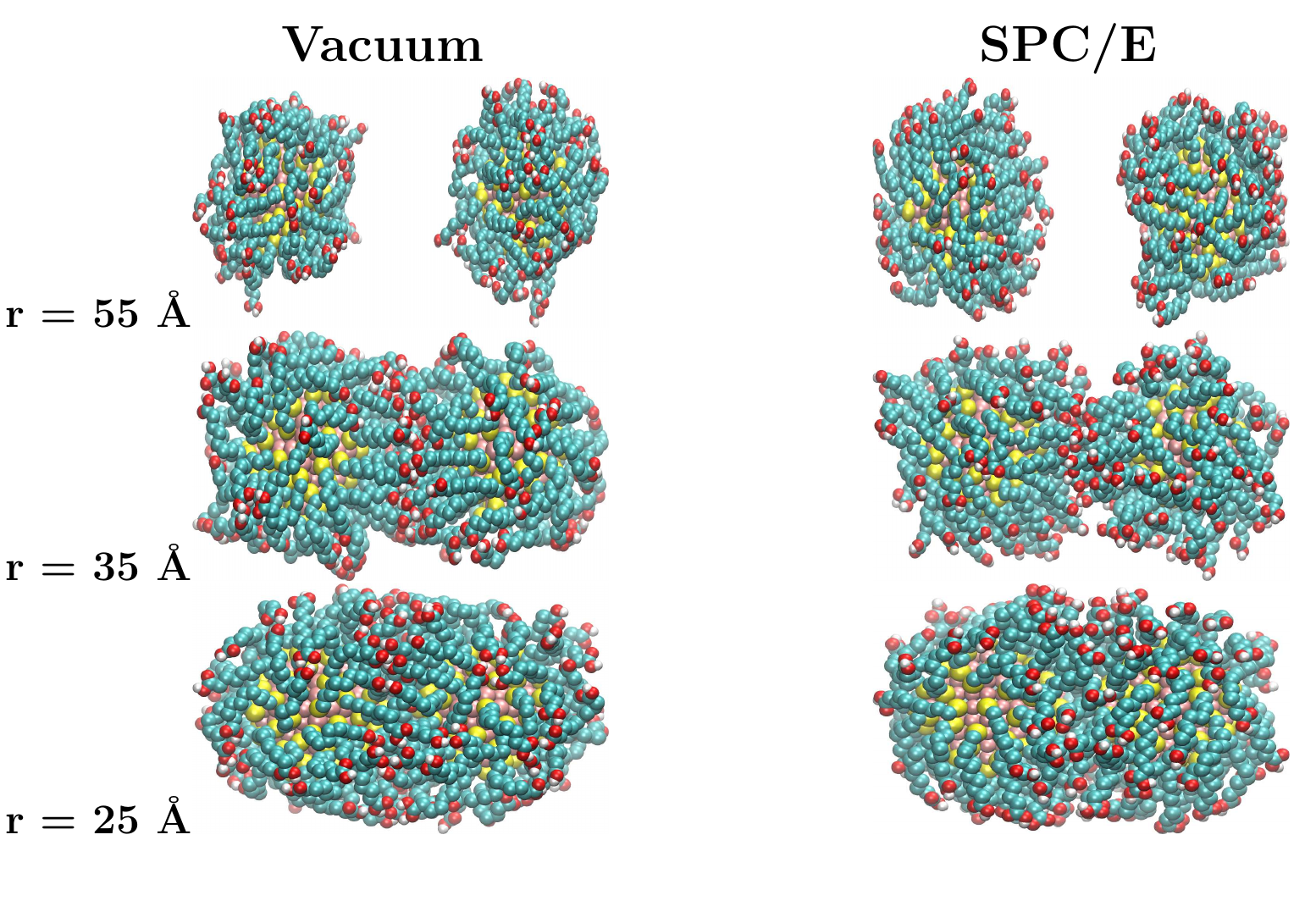}
    \end{center}
    \caption{Anisotropy due to ligand shell disorder at different pair
    separations for the \auc~nanoparticle in vacuum
  and SPC/E water at 300 K. The color coding is same as in
  Figure~\ref{fig:snap1}.}

        \label{fig:snap2}
\end{figure}

Figure~\ref{fig:snap2} shows snapshots of a pair of \auc~nanoparticles at
different inter-nanoparticle pair separations, $r$, in vacuum and SPC/E water. When the two
nanoparticles are at very large separation (55 \AA), the fluctuations in the
ligand shell give rise to inherent anisotropy. The inherent
anisotropy of all the nanoparticles including \auc~is more in SPC/E than in vacuum as can be seen clearly in
Figure~\ref{fig:rgr}(b). At $r$ = 55 \AA~distance, there is almost negligible interaction between the two NPs, but as the
pair separation is decreased, the interaction between the nanoparticles
increases,
giving rise to emergent anisotropy. 
In vacuum, due to the absence of ligand-solvent
interaction, the ligand-ligand interaction is more compared to SPC/E and hence
greater number of ligands align in the direction of the inter-nanoparticle axis compared to
SPC/E as can be seen at $r$ = 35 \AA ~in Figure~\ref{fig:snap2}. With the further
decrease in the pair separation, the ligands are pushed out from the space
between the two nanoparticles.

\begin{figure}[h!]
\begin{center}
  \includegraphics[clip=true,trim=4.4cm 9.7cm 4.3cm 7.9cm,width=16.0cm]{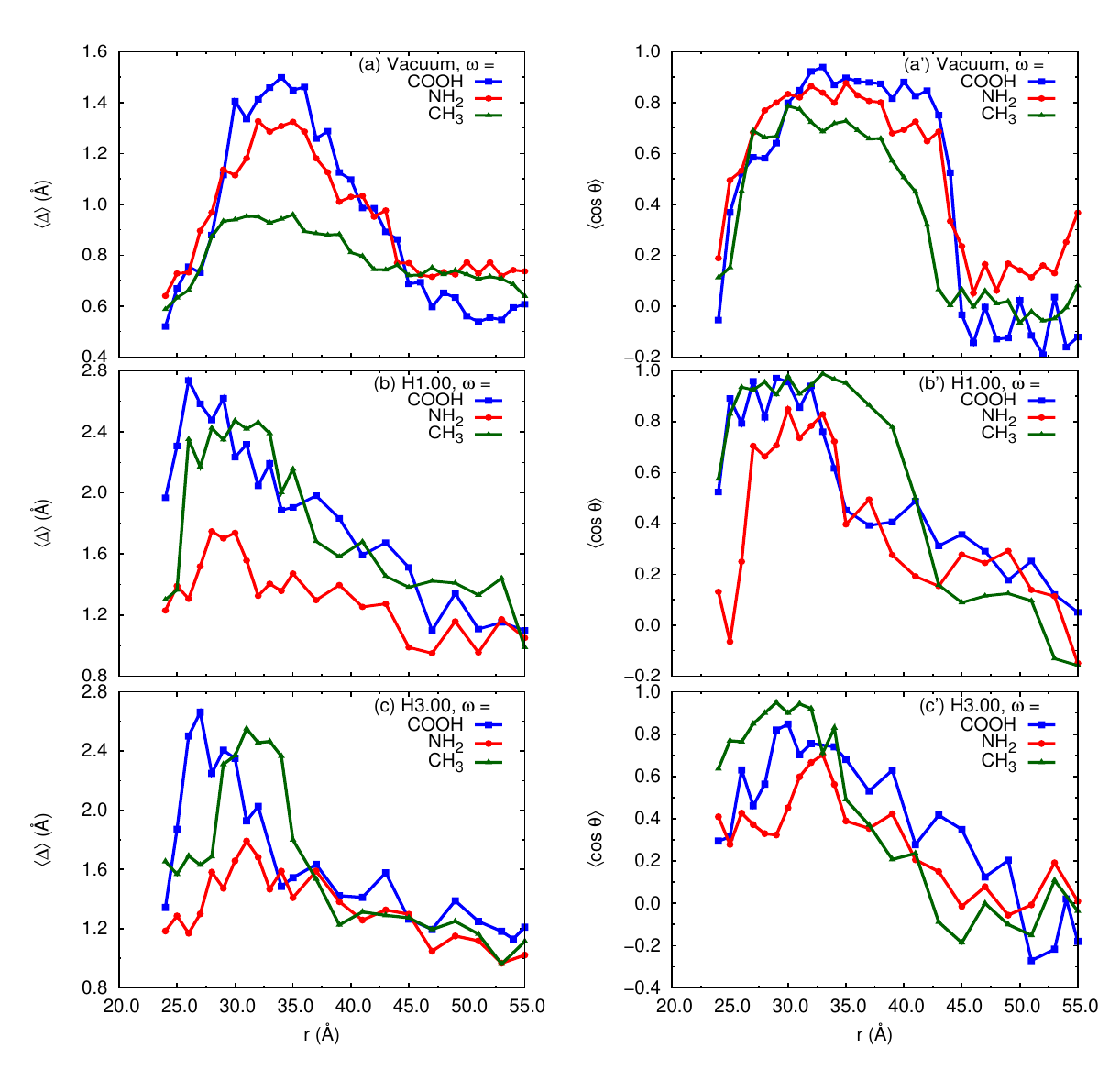}
    \end{center}
    \caption{Fluctuation-driven anisotropy in passivated nanoparticles as a
    function of pair separation, $r$, at 300 K. The mean value of the
  magnitude of the mass dipole, $\langle \Delta \rangle$, and the mean angle
between the mass dipole vector and the inter-nanoparticle axis, $\langle
\cos~\theta \rangle$, are shown in (a and a$^\prime$)
vacuum, (b and b$^\prime$) H1.00 or SPC/E and (c and c$^\prime$) H3.00, respectively.}
        \label{fig:delta}
\end{figure}

Figure~\ref{fig:delta} shows the change in the average mass dipole ($\langle \Delta
\rangle$) and the orientation of the mass dipole vector, characterized by the
cosine of the angle ($\cos \theta$)
between the mass dipole vector and the inter-nanoparticle axis with the change in
the pair separation between the two nanoparticles. The higher average mass dipole
value indicates higher asymmetry. The value of $\langle \cos~\theta \rangle$ can vary from -1 to 1.
The $\langle \Delta \rangle$ value for a pair of nanoparticles of all three types do not change much up to $\approx$ 45 \AA~ in vacuum, which
is close to $2R_c+2L$ value. With the further decrease in pair separation, the
$\langle \Delta \rangle$ increases up to $r \approx$ 33 \AA, which is close to
$2R_c+L$ value due to attractive energetic interactions between ligands as shown
by an attractive well in $V_{PMF}$ profiles (see Figure~\ref{fig:pmf}). With the further decrease in pair
separation, the $\langle \Delta \rangle$ decreases as the ligands are pushed out
of the inter-nanoparticle space due to the entropic repulsion of ligands as
characterized by the increase in the value of $V_{PMF}$. The 
increase is maximum for \auc~nanoparticle and minimum for \aum~nanoparticle. The change in the $\langle
\Delta \rangle$ is the result of the changes in the orientation of the ligands due
to decreasing pair separation between the nanoparticles. The $\langle \cos~\theta \rangle$
shows qualitatively similar behaviour as of the $\langle \Delta \rangle$. At
large separation, the $\langle \cos~\theta \rangle$ is close to zero and does not
change much up to $\approx$ 45 \AA, with further decrease in pair separation, the
$\langle \cos~\theta \rangle$ shows values close to 1.0 due to the
alignment of ligands in the same direction to the inter-nanoparticle axis. At very low pair 
separations, the ligands are pushed out from the
inter-nanoparticle space due to entropic repulsion between ligand atoms and the
$\langle \cos~\theta \rangle$ shows lower
values. 
In vacuum, the $\langle \Delta \rangle$ is relatively low at all pair
separations compared to $\langle \Delta \rangle$ values in SPC/E and H3.00 may
be due to higher inherent anisotropy observed for all single nanoparticles in these
solvents. In SPC/E and H3.00 solvents, the \aum~nanoparticle shows greater emergent
anisotropy compared to \aua~and \auc~nanoparticles at lower pair separations, which is in agreement with
the lower well depth value observed in $V_{PMF}$ profiles in these solvents for
\aum~nanoparticle. The emergent
anisotropy in solvents is maximum at very low pair separations, whereas in
vacuum, the emergent anisotropy is maximum at intermediate pair separations
between the two nanoparticles. 
\section{Conclusions}
\label{sec:conclu6}
This study examines the solvation behaviour of end group functionalized
alkylthiols passivated gold nanoparticles namely \aum, \aua~and \\
\auc~in solvents with varying degrees of
repulsion-dispersion and electrostatic interactions. 
The solvent reorganization around the end group functionalized nanoparticles are
characterized by the pair correlation functions, $g_{ns}(r)$, between the center of mass of the
gold nano-core and solvent atoms. The PCFs of functionalized nanoparticles in
SPC/E water show only slight differences, but the solvation properties
calculated using, $g_{ns}(r)$, show significant differences. As the
repulsion-dispersion interactions are enhanced relative to the electrostatic
interactions, the solvent penetrability inside the ligand soft corona increases
at lower $r$. Decane solvent shows maximum penetrability, even higher than H3.00
modified hybrid water. $g_{ns}(r) <$ 1.0, in the ligand soft corona for
combinations of all types of nanoparticles and solvents, due to greater
solvent-solvent interactions compared to solute-solvent interactions. The solvent
excess, $n^E_s$, is maximum (less negative) in decane and minimum (more
negative) in H1.00. It gives the quantitative measure of the affinity of solute
particle for a given solvent, hence the affinity of nanoparticles for decane
solvent is highest among all the solvents used in this study. 
At lower temperature, solute-solvent interaction is relatively high as indicated by less
negative value of $n^E_s$.
The local
entropy ($\Delta S^{loc}_{ns}$) and total entropy ($\Delta S^{tot}_{ns}$)
increase with increase in repulsion-dispersion interaction of modified hybrid
water models, but it is maximum in decane.

The changes in the local structure of water molecules in the presence of end group
functionalized nanoparticles is examined using the local
tetrahedral order metric, $q_{tet}$. At large $r$ from the center of mass of the
nano-core, the effect due to nanoparticles is negligible, even for \auc~nanoparticle.
At the periphery of the passivated nanoparticle, changes in the local structure
of water molecules are eminent. The change is maximum for \auc~nanoparticle as
evident from both $q_{tet}(r)$ and $P_r(q_{tet})$ plots. The degree of breaking
the tetrahedral structure of water molecules follow the order
\auc~$>$ \aua~$>$ \aum.

The structure of ligands are characterized using the radial density profiles,
the radius of gyration ($R_g$) and ligand asymmetry parameter ($\langle \Delta
\rangle$).
The ligands are more open and stretched in decane than water models,
probably due to more penetration of decane solvent inside the ligand shell. The
effect of functionalization is also more evident in decane than SPC/E water. The
$R_g$ values support the observations obtained from radial density profiles and
follow the order, decane $>$ vacuum $>$ H3.00 $\approx$ H1.56 $>$ SPC/E. The
highest anisotropy is seen in SPC/E and H3.00 water and least in decane. The
anisotropy is maximum for \auc~nanoparticle compared to \aum~and \aua~nanoparticles in all solvents, except 
in vacuum. The ligand shell configurational entropy ($S_L$) computed using the
covariances in particle displacement of ligand atoms for all three
nanoparticles follow the order, vacuum $>$ decane $>$ H1.56 $>$ H3.00 $\approx$
SPC/E. In all solvents, the $S_L$ follows the order \aum~$>$ \aua~$>$ \auc.

All the isotropic PMFs for identical nanoparticles in SPC/E and H3.00 water obtained in this study are
characterized by a deep attractive well similar to PMF profiles in vacuum. The
PMF profiles in SPC/E and H3.00 solvent are in contrast with the repulsive PMF profile of
\aum~nanoparticle in decane\cite{ysac16}. This indicates that SPC/E and H3.00 water at density
$\approx$ 1.0 g/cm$^3$ act as poor solvent conditions as compared to decane. All
the PMF profiles reported in this study are fitted with a Morse function. The well
depth, range parameter and the equilibrium separation are significantly
different for different nanoparticles in different solvents. In vacuum, the well
depth of \auc~nanoparticle is maximum and minimum for \aum~nanoparticle, but it
is reverse in SPC/E water. The well depths for all nanoparticles are minimum in
H3.00 water compared to vacuum and SPC/E water. The equilibrium separations,
$r_e$, are also quite different in vacuum, but the variations are less in solvent media. 
The variations arise due to differences in ligand-ligand energetic interaction and 
entropic effect due to ligands in different solvents. The average mass dipole
increases with the decrease in pair separation between nanoparticles and is
maximum in the region of attractive $V_{PMF}$. It decreases, when the ligands are pushed out of the inter-nanoparticle space due to entropic
repulsion between ligands. Similarly, the orientation of the mass dipole w.r.t
inter-nanoparticle axis, characterized by $\langle \cos \theta \rangle$, shows
value close to 1.0 in the region of attractive $V_{PMF}$, due to the alignment of ligands in the same direction to the
inter-nanoparticle axis. It is interesting to compare the emergent
anisotropy results obtained in this study with our previous study of \aum~
nanoparticle in decane. The $V_{PMF}$ for \aum~ in decane
solvent is repulsive in nature and in the repulsive regime the $\langle \Delta
\rangle$ increases significantly accompanied with a sharp decrease in $\langle
\cos~\theta \rangle$\cite{ysac16}. 

Our results indicate that the chemistry of the ligand and the solvent plays an
important role in controlling the solvation and aggregation behavior of
the nanoparticles. The structure of the coatings is also highly dependent on the
functionalization of passivating ligands and interactions present in the
solvent. The changes in the PMF profiles with increasing repulsion-dispersion
interaction of solvent signify the importance of solvent chemistry in
self-assembly process. It would be interesting to compare our PMF results with
PMF profiles for charged end group functionalized ligand passivated nanoparticles in
water.

{\bf{Acknowledgments}}

The authors are thankful to the Department of Science and Technology, India
for financial support. S.P. thanks the University Grants Commission
for support through a Senior Research Fellowship. Authors also thank the
IIT Delhi HPC facility for computational resources.

\clearpage
\bibliography{ref-thesis-saurav}
\end{document}


\setstretch{1.5}
\title{{\bf{Supplementary Material}}\\
Gold Nanoparticles Passivated with Functionalized Alkylthiols: Simulations of Solvation in the Infinite Dilution Limit}
\author{Saurav Prasad}
\email[Author for correspondence:]{ s.prasad@theo.chemie.tu-darmstadt.de}
\email[Tel:]{ +49-6151 16-22613}
\email[Fax:]{ +49-6151 16-22619.}
\affiliation{Eduard-Zintl-Institut f{\"u}r Anorganische und Physikalische Chemie, Technische Universit{\"a}t Darmstadt, Alarich-Weiss-Street 8, D-64287 Darmstadt, Germany.}
\author{Madhulika Gupta}
\affiliation{Department of Chemistry, Indian Institute of Technology Delhi, New Delhi:
110016, India.}

\date{\today}
\maketitle
\begin{table}[htbp]
  \caption{Potential energy parameters for hybrid modified water models\cite{ld05}. The
Lennard-Jones site is associated with the energy ($\epsilon_{OO}$) and length ($\sigma_{OO}$) parameters. The bond lengths and bond angles of the water molecules are denoted by $r_{OH}$ and $\angle HOH$, respectively. Partial charges on oxygen and hydrogen atoms are denoted by $q_O$ and $q_H$, respectively.}
\label{tab:pmw}
\centering
\setlength{\tabcolsep}{0.4cm}
\renewcommand{\arraystretch}{1.2}
\begin{tabular}{c c c c c c c}
             \hline
	     \hline
	     Model & $\epsilon_{OO} $ & $\sigma_{OO} $ &
             r$_{OH} $ & q$_{O} $ & q$_{H} $ & $\angle$HOH \\
             \hline 
             & (kJ/mol)  & ($\text{\AA}$) & ($\text{\AA}$) & (e) & (e) &
             ($^\circ$)\\
             \hline
	     SPC/E  & 0.6502 & 3.166 & 1.0 & -0.8472 & 0.4238 & 109.47 \\
	     H1.56  & 1.0143 & 3.166 & 1.0 & -0.8472 & 0.4238 & 109.47 \\
	     H3.00  & 1.9506 & 3.166 & 1.0 & -0.8472 & 0.4238 & 109.47 \\
	     \hline
	     \hline
\end{tabular}
\end{table}   

\begin{table}[htbp]
\caption{Lennard-Jones and Coulomb potential parameters for non-bonded interactions for the
$\omega$-functionalized gold nanoparticles;
$U_{ij}(r_{ij})=4\epsilon_{ij}[(\frac{\sigma_{ij}}{r_{ij}})^{12}-(\frac{\sigma_{ij}}{r_{ij}})^{6}]+
\frac{1}{4\pi\epsilon_{o}}\frac{q_iq_j}{r_{ij}}$. The $\epsilon$ and $\sigma$ are the
energy and size parameters respectively. Partial charges on different 
atoms are denoted by $q$. The letters in parentheses indicate the atom with which
the particular site is bonded.}
\label{tab:ljpar}
\centering
\setlength{\tabcolsep}{0.4cm}
\renewcommand{\arraystretch}{1.2}
\begin{tabular}{l r r r r}
             \hline
	     \hline
	     Atom pairs & $\epsilon $ & $\sigma$ & q & References \\
             \hline 
             & (kJ/mol)  & ($\text{\AA}$) & (e) & \\
             \hline
             CH$_2$ & 0.3825 & 3.950 & --- & \cite{ms98} \\
             CH$_3$ & 0.8148 & 3.750 & --- & \cite{ms98} \\
             SH     & 1.6629 & 4.450 & --- & \cite{tb05,ll04} \\
             Au     & 3.3388 & 2.737 & --- & \cite{tb05,ll04} \\
             N      & 0.9228 & 3.340 & -0.892 & \cite{rj99,wsrs05} \\
             H-(N)  & 0.0000 & 0.000 & +0.356 & \cite{rj99,wsrs05} \\
             CH$_2$-(N) & 0.3825 & 3.950 & +0.180 & \cite{rj99,wsrs05} \\
             C      & 0.3409 & 3.900 & +0.420 & \cite{kcp04,cbr06} \\
             O=(C)  & 0.6568 & 3.050 & -0.450 & \cite{kcp04,cbr06} \\
             O-(H)  & 0.7732 & 3.020 & -0.460 & \cite{kcp04,cbr06} \\
             H-(O)  & 0.0000 & 0.000 & +0.370 & \cite{kcp04,cbr06} \\
	     \hline
	     \hline
\end{tabular}
\end{table}  
\begin{table}[htbp]
\caption{Harmonic stretching potential parameters for the
$\omega$-functionalized gold nanoparticles; $U(r)=\frac{1}{2}k_r(r-r_0)^2$ where, $k_r$ is the force constant and
$r_0$ is the equilibrium bond length. Same references were used as given in
Table~\ref{tab:ljpar}.}
\label{tab:hars}
\centering
\setlength{\tabcolsep}{0.4cm}
\renewcommand{\arraystretch}{1.2}
\begin{tabular}{l c c}
             \hline
	     \hline
	     Bond type & $k_r$  & $r_0$ \\
             \hline 
             & (kJ/mol/\AA$^2$)  & ($\text{\AA}$)  \\
             \hline
             CH$_2$-CH$_x$        & 2250.00 & 1.5400 \\
             CH$_2$-CH$_2$-(N)    & 2250.00 & 1.5400 \\
             CH$_2$-CH$_2$-(C)    & 2250.00 & 1.5400 \\
             CH$_2$-SH  & 2250.00 & 1.8200 \\
             CH$_2$-N   & 3193.52 & 1.4480 \\
             N-H        & 3628.24 & 1.0100 \\
             C-O        & 1882.79 & 1.3640 \\
             CH$_2$-C   & 1326.32 & 1.5200 \\
             O-H        & 2313.74 & 0.9700 \\
             C=O        & 2384.87 & 1.2140 \\

	     \hline
	     \hline
\end{tabular}
\end{table}  
\begin{table}[htbp]
\caption{Harmonic bending potential parameters for the $\omega$-functionalized
gold nanoparticles;
$U(\theta)=\frac{1}{2}k_\theta(\theta-\theta_0)^2$ where, $k_\theta$ is the force
constant and $\theta_0$ is the equilibrium bond angle. Same references were used as given in
Table~\ref{tab:ljpar}.}
\label{tab:harb}
\centering
\setlength{\tabcolsep}{0.4cm}
\renewcommand{\arraystretch}{1.2}
\begin{tabular}{l c c}
             \hline
	     \hline
	     Angle type & $k_\theta$  & $\theta_0$ \\
             \hline 
             & (kJ/mol/rad$^2$)  & ($^\circ$)  \\
             \hline
             CH$_2$-CH$_2$-CH$_x$ & 519.6543 & 114.00 \\
             CH$_2$-CH$_2$-SH     & 519.6543 & 114.00 \\
             CH$_2$-CH$_2$-CH$_2$-(N)      & 519.6543 & 114.00 \\
             CH$_2$-CH$_2$-C      & 519.6543 & 114.00 \\
             CH$_2$-CH$_2$-N  & 470.5726 & 109.50 \\
             CH$_2$-N-H       & 519.6240 & 112.90 \\
             H-N-H            & 365.0000 & 106.40 \\
             CH$_2$-C-O       & 278.1946 & 111.00 \\
             CH$_2$-C=O       & 335.0728 & 126.00 \\
             O=C-O            & 335.0728 & 123.00 \\
             C-O-H            & 146.3345 & 107.00 \\
	     \hline
	     \hline
\end{tabular}
\end{table}
\begin{table}[htbp]
\caption{Triple cosine potential parameters for 1-4 torsional interaction in the
$\omega$-functionalized gold nanoparticles;
$U(\phi)=\frac{1}{2}A_1(1+cos(\phi))+\frac{1}{2}A_2(1-cos(2\phi))+\frac{1}{2}A_3(1+cos(3\phi))$
where A$_1$, A$_2$, and A$_3$ are constants. Same references were used as given in
Table~\ref{tab:ljpar}.}
\label{tab:trp}
\centering
\setlength{\tabcolsep}{0.4cm}
\renewcommand{\arraystretch}{1.2}
\begin{tabular}{l c c c}
             \hline
	     \hline
	     Dihedral type & A$_1$  & A$_2$ & A$_3$ \\
             \hline 
             & (kJ/mol)  & (kJ/mol) & (kJ/mol)  \\
             \hline
             CH$_2$-CH$_2$-CH$_2$-CH$_x$     & 5.9038 & -1.1339 & 13.1588 \\
             CH$_2$-CH$_2$-CH$_2$-CH$_2$-SH  & 5.9038 & -1.1339 & 13.1588 \\
             CH$_2$-CH$_2$-CH$_2$-CH$_2$-(N) & 5.9038 & -1.1339 & 13.1588 \\
             CH$_2$-CH$_2$-CH$_2$-CH$_2$-(C) & 5.9038 & -1.1339 & 13.1588 \\
             CH$_2$-CH$_2$-C-O               & 5.9038 & -1.1339 & 13.1588 \\
	     \hline
	     \hline
\end{tabular}
\end{table}
\begin{table}[htbp]
\caption{Multi-harmonic potential parameters for 1-4 torsional interaction in the
$\omega$-functionalized gold nanoparticles; $U(\phi)=\Sigma_{i=1}^{5}C_i
cos^{i-1}(\phi)$ where, $C_i$'s are constants. Same references were used as given in
Table~\ref{tab:ljpar}.}
\label{tab:mult}
\centering
\setlength{\tabcolsep}{0.4cm}
\renewcommand{\arraystretch}{1.2}
\begin{tabular}{l c c c c c}
             \hline
	     \hline
	     Dihedral type & C$_1$  & C$_2$ & C$_3$ & C$_4$ & C$_5$  \\
             \hline 
             & (kJ/mol)  & (kJ/mol) & (kJ/mol) & (kJ/mol) & (kJ/mol)  \\
             \hline
             CH$_2$-CH$_2$-CH$_2$-N          & 2.390  & 6.867 & 2.532 & -3.824 & -0.03822 \\
             CH$_2$-CH$_2$-N-H  & 0.7067 & 3.0160 & 1.7460 & -3.492 & -0.0000234 \\
             CH$_2$-CH$_2$-C=O  & 18.230 & -5.238 & -12.990 & -0.0000229 & -0.0000334 \\
             CH$_2$-C-O-H       & 18.230 &  5.238 & -12.990 & 0.0000373  & -0.0000362 \\
             O=C-O-H            & 18.230 & -5.238 & -12.990 & -0.0000229 & -0.0000334 \\
	     \hline
	     \hline
\end{tabular}
\end{table}
\begin{figure}[htbp]
\begin{center}
  \includegraphics[clip=true,trim=6.1cm 10cm 6.2cm 8.3cm,width=8.5cm]{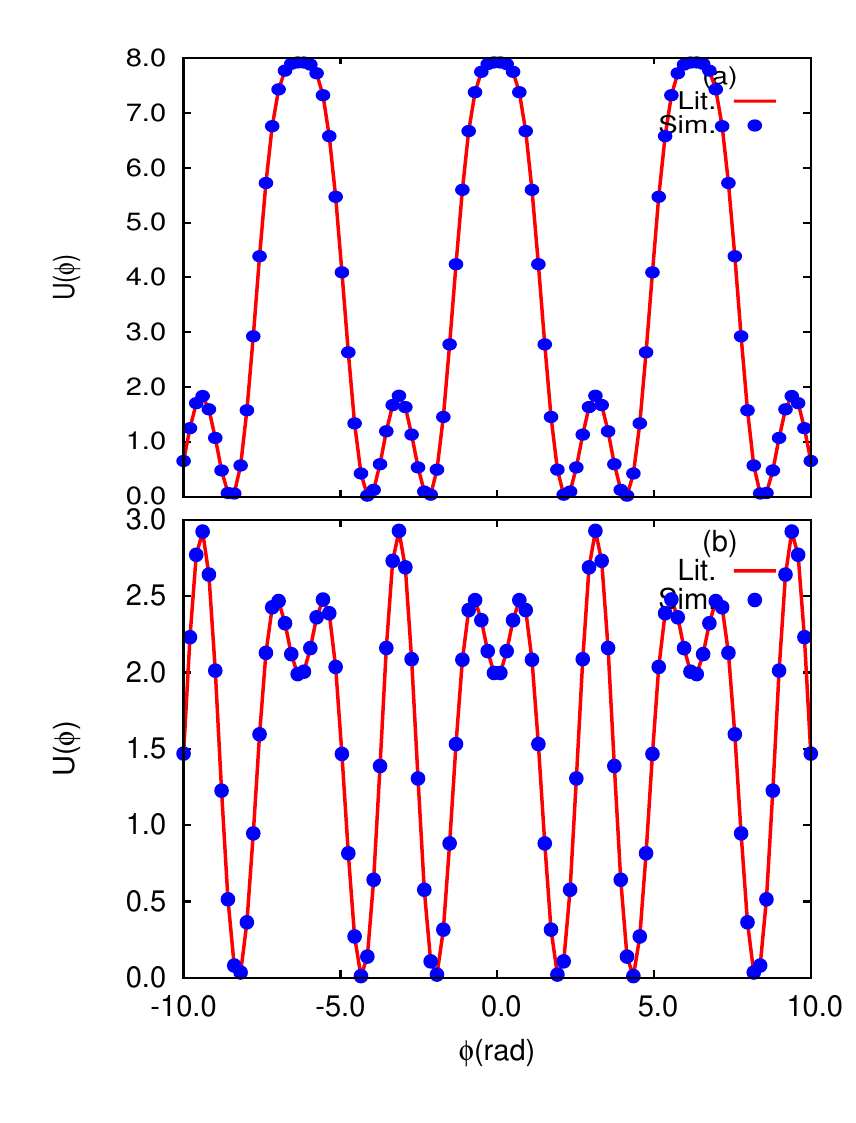}
    \end{center}
    \caption{Fitting of torsional potential used in literature and
    multi-harmonic torsional potential used in this study for (a)
  CH$_2$-CH$_2$-CH$_2$-N and (b) CH$_2$-CH$_2$-N-H dihedrals.}
        \label{fig:tor1}
\end{figure}
\begin{figure}[htbp]
\begin{center}
  \includegraphics[clip=true,trim=6.1cm 10cm 6.2cm 8.3cm,width=8.5cm]{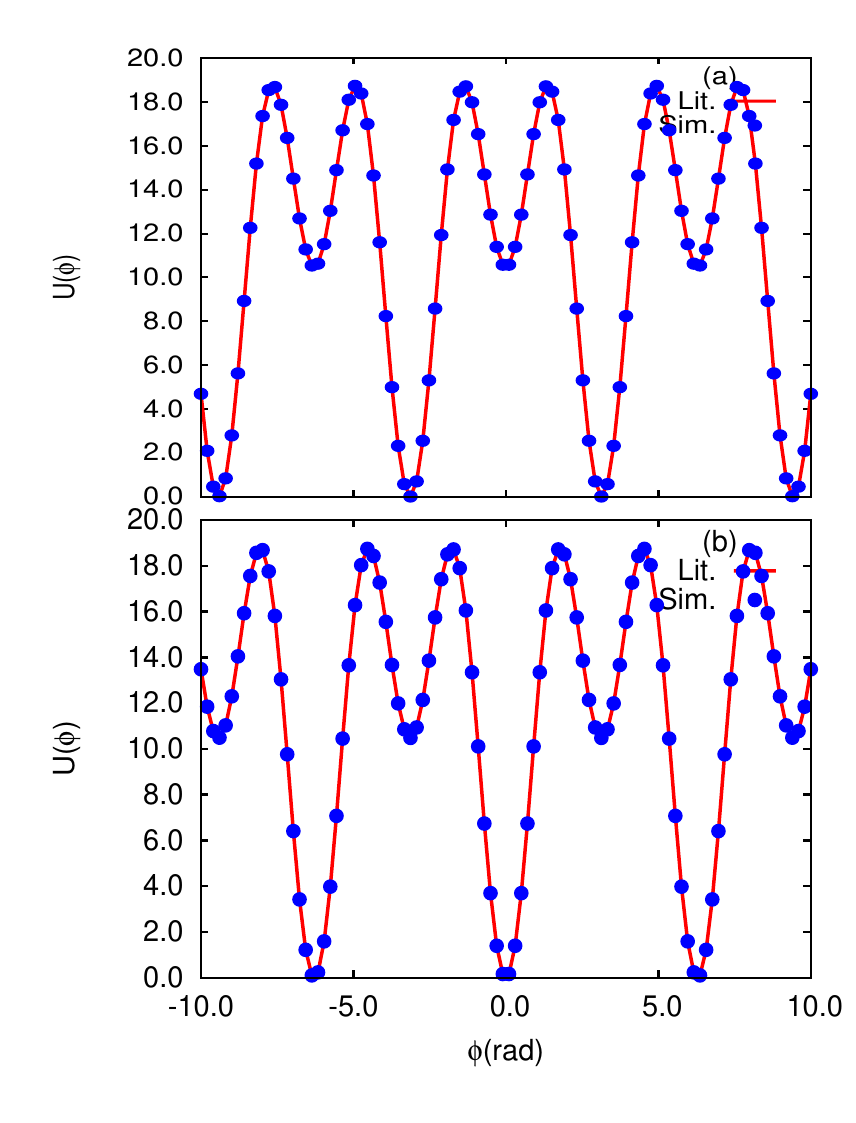}
    \end{center}
    \caption{Fitting of torsional potential used in literature and
        multi-harmonic torsional potential used in this study for (a) CH$_2$-C-O-H;
      (b) CH$_2$-CH$_2$-C=O and O=C-O-H dihedrals.}
        \label{fig:tor2}
\end{figure}

\begin{figure}[htbp]
\begin{center}
  \includegraphics[clip=true,height=5cm,width=14cm]{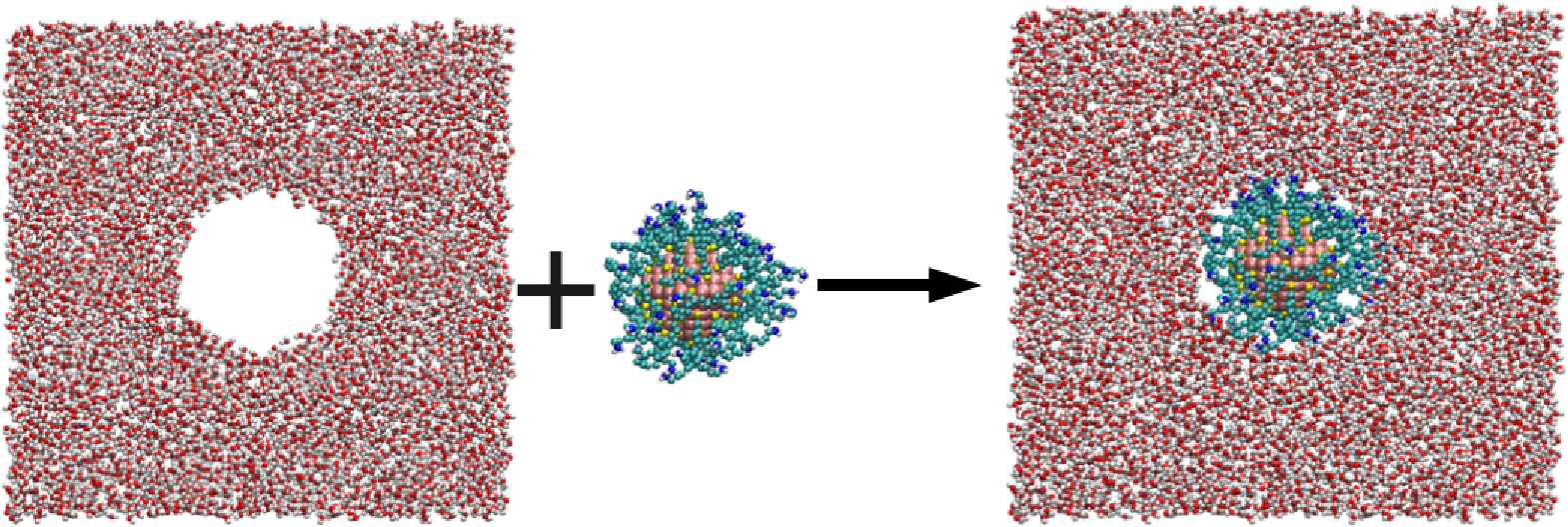}
    \end{center}
    \caption{Schematic diagram illustrating the insertion of \aua~nanoparticle
    in the solvent box.}
        \label{fig:solvate}
\end{figure}
\begin{figure}[htbp]
\begin{center}
  \includegraphics[clip=true,trim=6.1cm 10.0cm 6.2cm 8.4cm,width=8.5cm]{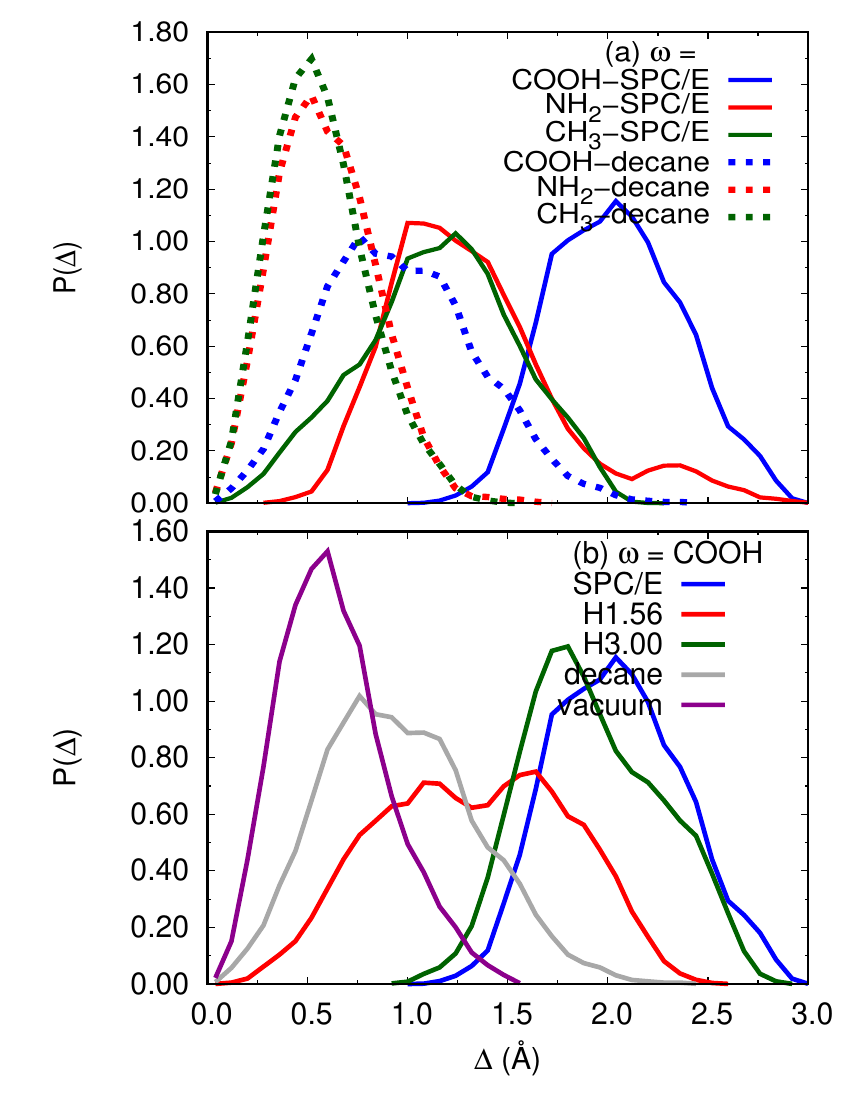}
    \end{center}
    \caption{The probability distribution, P($\Delta$): (a) for different functionalized nanoparticles in SPC/E water and decane at 300 K, 1 atm; (d) for COOH functionalized nanoparticle in various solvent media at 300 K, 1 atm.}
        \label{fig:fig5}
\end{figure}

\clearpage
\bibliography{ref-thesis-saurav}